\documentclass[onecolumn,amsmath,amssymb,superscriptaddress,a4paper,nofootinbib,accepted=2023-03-22]{quantumarticle}
\pdfoutput=1
\usepackage[utf8]{inputenc}
\usepackage[english]{babel}
\usepackage[T1]{fontenc}
\usepackage{graphicx,amsmath,color,bm,rotating,dsfont,stackrel,epstopdf,dsfont}

\usepackage{hyperref}
\usepackage[numbers]{natbib}

\usepackage{tikz}
\usepackage{lipsum}

\newcommand{\eq}[1]{\begin{equation}\begin{aligned}#1\end{aligned}\end{equation}}
\newcommand{\iu}{\text{i}}
\newcommand{\eu}{\text{e}}
\newcommand{\ha}{{a}}
\newcommand{\had}{{a}^\dagger\vphantom{a}}

\newcommand{\ket}[1]{\left\lvert#1\right\rangle}

\newcommand{\bra}[1]{\left\langle#1\right\rvert}

\newcommand{\braket}[2]{\left.\left\langle#1\right\rvert#2\right\rangle}

\newcommand{\expct}[1]{\left\langle#1\right\rangle}
\newcommand{\RE}{\mathop{\mathrm{Re}} \nolimits}
\newcommand{\IM}{\mathop{\mathrm{Im}} \nolimits}

\newcommand{\aaron}[1]{{#1}}

\usepackage{soul,xcolor}

\DeclareMathOperator{\sinc}{sinc}

\begin{document}

\title{Beyond transcoherent states: Field states for effecting optimal coherent rotations on single or multiple qubits}
\author{Aaron Z. Goldberg}
\affiliation{National Research Council of Canada, 100 Sussex Drive, Ottawa, Ontario K1N 5A2, Canada}
%\affiliation{Department of Physics and Centre for Quantum Information \& Quantum Control, University of Toronto, Toronto, Ontario, Canada M5S 1A7}
\affiliation{Department of Physics, University of Ottawa, 25 Templeton Street, Ottawa, Ontario, K1N 6N5 Canada}
\author{Aephraim M. Steinberg}
\affiliation{Department of Physics and Centre for Quantum Information \& Quantum Control, University of Toronto, Toronto, Ontario, Canada M5S 1A7}
\affiliation{CIFAR, 661 University Ave., Toronto, Ontario M5G 1M1, Canada}
\author{Khabat Heshami}
\affiliation{National Research Council of Canada, 100 Sussex Drive, Ottawa, Ontario K1N 5A2, Canada}
\affiliation{Department of Physics, University of Ottawa, 25 Templeton Street, Ottawa, Ontario, K1N 6N5 Canada}
\affiliation{Institute for Quantum Science and Technology, Department of Physics and Astronomy, University of Calgary, Alberta T2N 1N4, Canada}

\maketitle

\begin{abstract}
    Semiclassically,  laser pulses can be used to implement arbitrary transformations on atomic systems; quantum mechanically, residual atom-field entanglement spoils this promise. Transcoherent states are field states that fix this problem in the fully quantized regime by generating perfect coherence in an atom initially in its ground or excited state. We extend this fully quantized paradigm in four directions: First, we introduce field states that transform an atom from its ground or excited state to any point on the Bloch sphere without residual atom-field entanglement. The best strong pulses for carrying out rotations by angle $\theta$ are are squeezed in photon-number variance by a factor of $\sinc\theta$. Next, we investigate implementing rotation gates, showing that the optimal Gaussian field state for enacting a $\theta$ pulse on an atom in an arbitrary, unknown initial state is number squeezed by less: $\sinc\tfrac{\theta}{2}$. Third, we extend these investigations to fields interacting with multiple atoms simultaneously, discovering once again that number squeezing by $\tfrac{\pi}{2}$ is optimal for enacting $\tfrac{\pi}{2}$ pulses on all of the atoms simultaneously, with small corrections on the order of the ratio of the number of atoms to the average number of photons. Finally, we find field states that best perform arbitrary rotations by $\theta$ through nonlinear interactions involving $m$-photon absorption, where the same optimal squeezing factor is found to be $\sinc\theta$. Backaction in a wide variety of atom-field interactions can thus be mitigated by squeezing the control fields by optimal amounts.
\end{abstract}

\tableofcontents

\section{Introduction}~
Coherence underlies quantum phenomena. Familiar from waves, coherence gives rise to interference effects that power quantum systems to be different from and more useful than everyday objects. Quantum coherence enables computation \cite{Shor1999}, measurement \cite{Dowling1998}, teleportation \cite{Bouwmeesteretal1997}, and more, making it an important resource to quantify \cite{Aberg2006arxiv,Baumgratzetal2014,LeviMintert2014,WinterYang2016}. Our ability to generate and transform coherence is thus vital to the success of these ventures.
Here, we show how to ideally transfer arbitrary amounts of coherence from light to atomic systems.

Previous work found the ideal states of light for transferring maximal coherence to a single atom: \textit{transcoherent} states do this job and can be approximated by easier-to-generate squeezed light in the appropriate limits \cite{GoldbergSteinberg2020}. These are important to the plethora of applications requiring maximally coherent atomic states, such as quantum engines \cite{Korzekwaetal2016} and quantum state preparation with quantum logic gates \cite{Mulleretal2009}. In other applications, arbitrary superpositions of a two-level atom's ground and excited states $\ket{\mathrm{g}}$ and $\ket{\mathrm{e}}$ may be desired, with the most general state being
\eq{
\ket{\theta,\phi}\equiv \cos\frac{\theta}{2}\ket{\mathrm{g}}+\sin\frac{\theta}{2}\eu^{\iu\phi}\ket{\mathrm{e}};
\label{eq:theta phi atomic state}
} the atom may be a physical atom or any other physical system with two energy levels, known as a qubit. We find the ideal states for all of these applications and demonstrate how they avoid residual light-atom entanglement and other deleterious effects that ruin the quality of the atomic coherence.

Light is routinely used for controlling atomic states. Strong, classical light with a frequency close to the transition frequency of a two-level atom induces ``Rabi flopping'' that coherently drives the atom between $\ket{\mathrm{g}}$ and $\ket{\mathrm{e}}$ at the Rabi frequency $\Omega_0\sqrt{\bar{n}}$, where  $\Omega_0$ is the single-excitation Rabi frequency [sometimes known as the vacuum Rabi frequency, taking values from
kHz to tens of MHz in atomic systems, depending on cavity parameters, and up to hundreds of MHz for circuit quantum electrodynamics (QED) systems] and $\bar{n}$ is equal to the intensity of the field in the appropriate units that amount to the single-photon intensity when the field is quantized \cite{AllenEberly1987}. Waiting an appropriate time $\Omega_0\sqrt{\bar{n}}t=\theta$, for example, will lead to an atom in state $\ket{\mathrm{g}}$ rotating to state $\ket{\theta,0}$. However, even quasiclassical light in a coherent state is fundamentally made from a superposition of different numbers of photons \cite{Sudarshan1963,Glauber1963}, which each drive oscillations in the atom at a different Rabi frequency; these give rise to famous effects such as the collapses and revivals of Rabi oscillations that help demonstrate the existence of quantized photons underlying quasiclassical light \cite{Eberlyetal1980,Rempeetal1987}.

When considering the quantized version of light's interaction with a single atom, the Jaynes-Cummings model (JCM) dictates that the light will generally become entangled with the atom \cite{GeaBanacloche1990,GeaBanacloche1991,PhoenixKnight1991a,GeaBanacloche2002,vanEnkKimble2002,SilberfarbDeutsch}. This prevents the atom from being in any pure state $\ket{\theta,\phi}$ and always tends to degrade the quality of the atomic state thus created. Such is the problem that transcoherent states surmount for $\theta=\tfrac{\pi}{2}$ and that we generalize here.

A natural, further generalization of these results is to field states interacting with a collection of atoms. While the dynamics between a single atom and a mode of light are straightforward to solve through the JCM, the same with a collection of atoms, known as the Tavis-Cummings model (TCM), cannot usually be done in closed form. This enriches our problem and allows us to incorporate strategies from semiclassical quantization in\aaron{to} our investigations.

The above interactions are linear in the electromagnetic field operators. We lastly extend these results for arbitrary atomic control to interactions that involve nonlinear contributions from the electromagnetic field, such as $m$-photon absorption processes. These showcase the reach of our transcoherence idea well beyond the initial goal of transferring coherence from light to atoms.

\subsection{Jaynes-Cummings model}~
The Jaynes-Cummings Hamiltonian governs the resonant interaction between a single bosonic mode annihilated by $\ha$ and a two-level atom with ground and excited states $\ket{\mathrm{g}}$ and $\ket{\mathrm{e}}$, respectively:
\eq{
H=\omega\left(\had\ha+\ket{\mathrm{e}}\bra{\mathrm{e}}\right)+\frac{\Omega_0}{2}\left(\ha\sigma_+ + \had\sigma_-\right),
} where $\omega$ is the resonance frequency (ranging from hundreds of THz for optical transitions in atoms down to tens of GHz for transitions between Rydberg states to below ten GHz for hyperfine transitions or superconducting qubits), $\sigma_+=\sigma_-^\dagger=\ket{\mathrm{e}}\bra{\mathrm{g}}$ is the atomic raising operator, and we employ units with $\hbar=1$ throughout. The JCM characterizes light-matter interactions in a variety of physical systems including circuit QED \cite{Raimondetal2001}, cavity QED \cite{Finketal2008}, and parametric amplification \cite{GutierrezJaureguiAgarwal2021}. This interaction conserves total energy and total excitation number, as can be seen from its eigenstates
\eq{
\ket{\pm,n}=\frac{\ket{n}\otimes\ket{\mathrm{e}}\pm\ket{n+1}\otimes\ket{\mathrm{g}}}{\sqrt{2}},
\label{eq:JCM eigenstates}
} with eigenenergies $\pm\tfrac{\Omega_n}{2}$ for the quantized Rabi frequencies 
\eq{
\Omega_n=\Omega_0\sqrt{n+1}.
\label{eq:omega n definition}
} These are responsible for the field and the atom periodically exchanging an excitation with frequency $\tfrac{\Omega_n}{2}$ when the initial state is either $\ket{n}\otimes\ket{\mathrm{e}}$ or $\ket{n+1}\otimes\ket{\mathrm{g}}$.
We will work in the interaction picture with Hamiltonian
\eq{
H_{\mathrm{I}}=
\frac{\Omega_0}{2}\left(\ha\sigma_+ + \had\sigma_-\right);
} the Schr\"odinger-picture results can thence be obtained with the substitutions $\ket{n}\to\eu^{-\iu\omega nt}\ket{n}$ and $\ket{\mathrm{e}}\to\eu^{-\iu \omega t}\ket{\mathrm{e}}$.

When the atom is initially in its ground state and the field in state $\sum_n \psi_n\ket{n}$, the evolved state takes the form
\eq{
\ket{\Psi(t)}=&\psi_0\ket{0}\otimes\ket{\mathrm{g}}+\sum_{n=0}^\infty \psi_{n+1}\left(\cos\frac{\Omega_n t}{2}\ket{n+1}\otimes\ket{\mathrm{g}}
%\right.\\&\qquad\qquad\qquad\qquad\left.
-\iu\sin\frac{\Omega_n t}{2}\ket{n}\otimes \ket{\mathrm{e}}\right)\\
=&\sum_{n=0}^\infty \ket{n}\otimes\left(\psi_n\cos\frac{\Omega_{n-1}t}{2}\ket{\mathrm{g}}-\iu \psi_{n+1}\sin\frac{\Omega_n t}{2}\ket{\mathrm{e}}\right).
\label{eq:JCM from ground}
} Similarly,
when the atom is initially in its excited state and the field in state $\sum_n \psi_n\ket{n}$, the evolved state takes the form
\eq{
\ket{\Psi(t)}=&\sum_{n=0}^\infty \ket{n}\otimes \left(\psi_n\cos\frac{\Omega_n t}{2}\ket{\mathrm{e}}
-\iu \psi_{n-1}\sin\frac{\Omega_{n-1} t}{2}\ket{\mathrm{g}}\right),
\label{eq:JCM from excited}
} \aaron{where we employ a slight abuse of notation whereby  $\psi_{-1}=0$ because photon numbers must be positive and $\Omega_{-1}=0$ by extending the definition from Eq. \eqref{eq:omega n definition}.} 

To achieve an arbitrary final atomic state, the most intuitive procedure is to begin with the atom in its ground state and the field in the target atomic state $\cos\frac{\theta}{2}\ket{0}+\iu\sin\frac{\theta}{2}\eu^{\iu\phi}\ket{1}$ and wait for the duration of a ``single-excitation $\pi$ pulse'' $\Omega_0 t=\pi$ to enact the transformation
\eq{
&\left(\cos\frac{\theta}{2}\ket{0}+\iu\sin\frac{\theta}{2}\eu^{\iu\phi}\ket{1}\right)\otimes\ket{\mathrm{g}}
%\\
%&\qquad\qquad
\to\ket{0}
\otimes \left(\cos\frac{\theta}{2}\ket{\mathrm{g}}+\sin\frac{\theta}{2}\eu^{\iu\phi}\ket{\mathrm{e}}\right)= \ket{0}\otimes\ket{\theta,\phi}.
} We exhaustively show in Section \ref{sec:perfectly generating arbitrary coherence} how to achieve this transformation with other field states, with no residual atom-field entanglement, at faster rates, and with more feasible pulses of light.
Since the free atomic evolution enacts $\phi\to\phi-\omega t$, we can generate the states $\ket{\theta,\phi}$ with any value of $\phi$ and simply allow free evolution to generate the same state with any other value of $\phi$, so in the following we set $\phi=0$ (alternatively, direct solutions with $\phi\neq 0$ can readily be obtained by adjusting the relative phases between the photon-number states).

\section{Optimal field states for generating arbitrary amounts of atomic coherence}~
\label{sec:perfectly generating arbitrary coherence}
What are the optimal field states that can generate arbitrary pulse areas? That is, which field states
\eq{
\ket{\psi}=\sum_n \psi_n\ket{n}
} can achieve the transformations
\eq{
\ket{\psi}\otimes\ket{\mathrm{g}}\to\ket{\psi^\prime}\otimes \ket{\theta}
\qquad\mathrm{or}\qquad
\ket{\psi}\otimes\ket{\mathrm{e}}\to \ket{\psi^\prime}\otimes \ket{\theta},
} where the former correspond to ``$\theta$ pulses,'' the latter to ``$\theta+\pi$ pulses,'' and we have defined the atomic state $\ket{\theta}\equiv\ket{\theta,0}$ by allowing $\theta$ to extend to $2\pi$?
%\eq{
%\ket{\theta}\equiv\ket{\theta,0}=\cos\frac{\theta}{2}\ket{\mathrm{g}}+\sin\frac{\theta}{2}\eu^{\iu\phi}\ket{\mathrm{e}}?
%}  
We specifically seek transformations for which the final state has zero residual entanglement between the atom and the light, such that the atomic state can be used in arbitrary quantum information protocols without degradation.

\subsection{Transcoherent states}~
In Ref. \cite{GoldbergSteinberg2020}\aaron{,} we defined \textit{transcoherent states} as those enabling $\tfrac{\pi}{2}$ pulses. For atoms initially in their ground states, perfect $\tfrac{\pi}{2}$ pulses can be achieved by the transcoherent states whose coefficients in the photon-number basis satisfy the recursion relation
\eq{
\psi_{n+1}=\iu\frac{\cos\frac{\Omega_{n-1} t}{2}}{\sin\frac{\Omega_n t}{2}}\psi_n
\label{eq:recurrence relation transcoherent ground}
} to ensure that the amplitudes of $\ket{\mathrm{g}}$ and $\ket{\mathrm{e}}$ in the evolved state in Eq. \eqref{eq:JCM from ground} are equal. This can be satisfied by field states with $\psi_n=0$ for $n> n_{\mathrm{max}}$ for \textit{any} chosen maximum photon number $n_{\mathrm{max}}\geq 1$, so long as the total interaction time satisfies
\eq{
\Omega_{n_{\mathrm{max}}-1}t=\pi,
\label{eq:interaction time transcoherent ground}
} which ensures that the highest-excitation subspace spanned by $\ket{\pm,n_{\mathrm{max}}-1}$ undergoes a $\pi$ pulse such that it transfers all of its excitation probability from $\ket{n_{\mathrm{max}}}\otimes\ket{\mathrm{g}}$ to $\ket{n_{\mathrm{max}}-1}\otimes\ket{\mathrm{e}}$. 
In the large-$n_{\mathrm{max}}$ limit, these states strongly approximate Gaussian states with an average photon number $\bar{n}$ whose photon-number distributions are squeezed from that of a canonical coherent state, $\sigma^2_{\mathrm{coh}}=\bar{n}$, by a factor of $\tfrac{\pi}{2}$.

Similarly, another set of transcoherent states has its photon-number distribution satisfy the same recursion relation as Eq. \eqref{eq:recurrence relation transcoherent ground}, with the lowest-excitation manifold undergoing a $(2k)\pi$ pulse and the highest a $(2k+1)\pi$ pulse for any $k\in\mathds{N}_0$. This pulse, in the large-$\bar{n}$ limit, corresponds to a $\frac{4k+1}{2}\pi$ pulse produced by a coherent state with its photon-number distribution squeezed by a factor of $\frac{4k+1}{2}\pi$.
Superpositions of such states with nonzero coefficients all satisfying $(2k)^2 n_{\mathrm{max}}\leq n\leq (2k+1)^2 n_{\mathrm{max}}$ will also enact perfect $\tfrac{\pi}{2}$ pulses in a time $\Omega_{n_{\mathrm{max}}-1}t=\pi$.

Another set of transcoherent states is found when the atom is initially in its excited state. This setup requires the initial field state's coefficients to satisfy a different recursion relation to ensure equal coefficients of $\ket{\mathrm{g}}$ and $\ket{\mathrm{e}}$ in Eq. \eqref{eq:JCM from excited}:
\eq{
 \psi_{n+1}=-\iu\frac{\sin\frac{\Omega_{n} t}{2}}{\cos\frac{\Omega_{n+1} t}{2}}\psi_{n}.
\label{eq:recurrence relation transcoherent excited} 
} As well, the lowest-excitation sector must now undergo a $(2k+1)\pi$ pulse while the highest undergoes a $2(2k+1)\pi$ pulse. This happens in a time \eq{
\Omega_{n_{\mathrm{min}}}t=(2k+1)\pi,
\label{eq:interaction time transcoherent excited}
} corresponding in the large-$\bar{n}$ limit to a $\tfrac{4k+3}{2}\pi$ pulse generated by a coherent state that has been photon-number squeezed by $\tfrac{4k+3}{2}\pi$. Superpositions of commensurate states will again achieve perfect coherence transfer.

\subsection{Beyond transcoherent states}~
We can generalize the recursion relations of Eqs. \eqref{eq:recurrence relation transcoherent ground} and \eqref{eq:recurrence relation transcoherent excited} to generate arbitrary \aaron{atomic} states of the form of Eq. \eqref{eq:theta phi atomic state}.

\begin{figure}
    \centering
    \includegraphics[width=0.65\columnwidth]{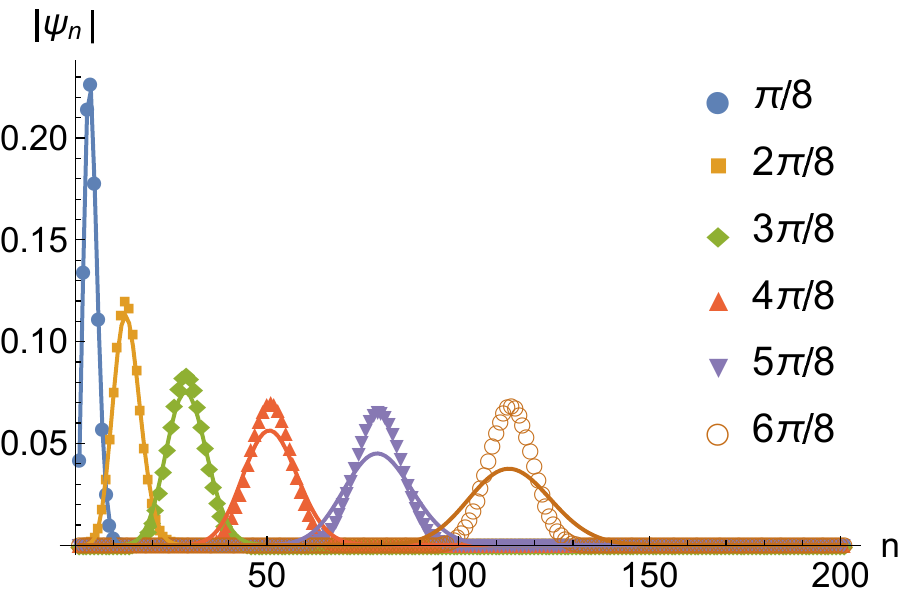}
    \caption{Photon-number probability distributions for field states that exactly generate arbitrary rotations $\theta$ on atoms initially in their ground states (various shapes correspond to different values of $\theta$). The field states are calculated using the recursion relation Eq. \eqref{eq:recurrence relation beyond transcoherent ground} with $n_{\mathrm{max}}=200$. Also plotted are the photon-number distributions for coherent states with the same average energies (solid curves). For the same $n_{\mathrm{max}}$ and thus the same value of $\Omega_0 t$, a higher-energy pulse generates a larger rotation angle $\theta$, with more photon-number squeezing being necessary for larger rotation angles.}
    \label{fig:beyond transcoherent from g}
\end{figure}

When the atom is initially in its ground state, it will evolve to a state of the form of Eq. \eqref{eq:theta phi atomic state} if and only if the initial field state's coefficients obey the recursion relation [again, c.f. Eq. \eqref{eq:JCM from ground}]
\eq{
\psi_{n+1}=\iu\tan{\frac{\theta}{2}}\frac{\cos\frac{\Omega_{n-1} t}{2}}{\sin\frac{\Omega_n t}{2}}\psi_n.
\label{eq:recurrence relation beyond transcoherent ground}
} The same boundary conditions as for transcoherent states hold, meaning that we require interaction times of the form of Eq. \eqref{eq:interaction time transcoherent ground} such that the lowest-excitation manifold undergoes a $0\pi$ pulse and the highest a $\pi$ pulse; extensions to other excitation manifolds are similarly possible. We plot a number of such states in Fig. \ref{fig:beyond transcoherent from g}.

When the atom is initially in its excited state, it will evolve to a state of the form of Eq. \eqref{eq:theta phi atomic state} if and only if the initial field state's coefficients obey the recursion relation [again, c.f. Eq. \eqref{eq:JCM from excited}]
\eq{
\psi_{n+1}=-\iu\tan{\frac{\theta}{2}}\frac{\sin\frac{\Omega_{n} t}{2}}{\cos\frac{\Omega_{n+1} t}{2}}\psi_{n}
.
\label{eq:recurrence relation beyond transcoherent excited}
} The same boundary conditions as for transcoherent states hold, meaning that we require interaction times of the form of Eq. \eqref{eq:interaction time transcoherent excited} such that the lowest-excitation manifold undergoes a $(2k+1)\pi$ pulse and the highest a $(4k+2)\pi$ pulse; extensions to superpositions of excitation manifolds are again possible. 

What are the properties of these extended transcoherent states whose coefficients are given\aaron{,} respectively\aaron{,} by Eqs. \eqref{eq:recurrence relation beyond transcoherent ground} and \eqref{eq:recurrence relation beyond transcoherent excited}? In Ref. \cite{GoldbergSteinberg2020}, we discussed how transcoherent states maximize the coherence \eq{
\mathcal{C}(t)&=\left|\bra{\Psi(t)}\sigma_+\ket{\Psi(t)}\right|+\left|\bra{\Psi(t)}\sigma_-\ket{\Psi(t)}\right|
\aaron{\propto \left|\sum_{n=0}^\infty \psi_{n+1}^*\psi_n\sin\frac{\Omega_n t}{2}\cos\frac{\Omega_{n-1}t}{2}\right|}.
\label{eq:coherence definition}
}  This is achieved through a careful balance between the narrowness of the distribution $|\psi_n|^2$, which favours nearly equivalent Rabi frequencies \aaron{such that the distribution $|\psi_n|^2$ is highly concentrated near the value of $n$ where the factor $\sin\frac{\Omega_n t}{2}\cos\frac{\Omega_{n-1}t}{2}$ from Eq. \eqref{eq:coherence definition} is peaked}, and its broadness, which leads to larger overlap terms $|\psi_{n+1}\psi_n|$ in Eq. 
\eqref{eq:coherence definition}. 
By creating the atomic states of Eq. \eqref{eq:theta phi atomic state}, we are attaining arbitrary values of $\mathcal{C}(t)$. How do the conditions of Eqs. \eqref{eq:recurrence relation beyond transcoherent ground} and \eqref{eq:recurrence relation beyond transcoherent excited} achieve the optimal balance for the distribution $|\psi_n|^2$?

We begin with the case of atoms initially in their ground states: Eq. \eqref{eq:recurrence relation beyond transcoherent ground}. By choosing states that satisfy the condition of Eq. \eqref{eq:interaction time transcoherent ground}, we ensure that the recursion relation truncates. Then, since the ratio $|\psi_{n+1}/\psi_n|$ monotonically decreases until it reaches zero, where the series truncates, the photon-number distribution approaches a smooth, singly peaked distribution;
for sufficiently large $\bar{n}$, this distribution is Gaussian.

The recursion relation of Eq. \eqref{eq:recurrence relation beyond transcoherent ground} is stationary when \eq{
\cos\frac{\theta}{2}\sin\frac{\Omega_n t}{2}-\sin\frac{\theta}{2}\cos\frac{\Omega_{n-1}t}{2}=0.
} For large $\bar{n}$, this occurs at a sufficiently large value of $n$ such that $\Omega_n\approx \Omega_{n-1}$; we will refer to this value as $\tilde{n}$, which we will later see to be on the order of the average number of photons $\tilde{n}=\mathcal{O}(\bar{n})$. This leads to the condition
\eq{
\sin\left(\frac{\theta}{2}-\frac{\Omega_{\tilde{n}} t}{2}\right)\approx 0\qquad
\Rightarrow\qquad \Omega_{\tilde{n}} t\approx \theta ,
\label{eq:stationary point ground}
}
corresponding with the classical scenario in which a pulse area of $\theta$ is applied when the Rabi frequency and total interaction time satisfy $\Omega_{\tilde{n}} t=\theta$.

Substituting this condition into  Eq. \eqref{eq:recurrence relation beyond transcoherent ground} leads to the expansion about large $\tilde{n}$:
\eq{
\tan{\frac{\theta}{2}}\frac{\cos\frac{\Omega_{{n}-1} t}{2}}{\sin\frac{\Omega_{{n}} t}{2}}=&\tan{\frac{\theta}{2}}\frac{\cos\frac{\theta\sqrt{n}}{2\sqrt{\tilde{n}+1}}}{\sin\frac{\theta\sqrt{n+1}}{2\sqrt{\tilde{n}+1}}}=\tan{\frac{\theta}{2}}\frac{\cos\frac{\theta\sqrt{\tilde{n}+\delta}}{2\sqrt{\tilde{n}+1}}}{\sin\frac{\theta\sqrt{\tilde{n}+\delta+1}}{2\sqrt{\tilde{n}+1}}}%\\
%&
\approx 1- \frac{\theta}{2\tilde{n}\sin\theta}\left(\delta-\sin^2\frac{\theta}{2}\right)
.} From this we find the approximate difference relation:
\eq{
\frac{\psi_{\tilde{n}+\delta}-\psi_{\tilde{n}}}{\delta}\approx -\frac{\theta}{2\tilde{n}\sin\theta}\left(\delta-\sin^2\frac{\theta}{2}\right)\psi_{\tilde{n}}.
} This describes a Gaussian distribution
\eq{
|\psi_{\tilde{n}+\delta}|^2\approx |\psi_{\tilde{n}}|^2 \exp\left[
-\frac{\theta}{2\tilde{n}\sin\theta}\left(\delta-\sin^2\frac{\theta}{2}\right)^2
\right]
\label{eq:Gaussian fround ground}
} with photon-number variance
\eq{
\sigma^2=\tilde{n}\sinc\theta.
\label{eq:sinc variance}
} The mean is slightly shifted to $\bar{n}=\tilde{n}+\sin^2\frac{\theta}{2}$, due to the discreteness of $n$; had we instead set the stationary point of the recursion relation to be at $\tilde{n}-1$, we would have accordingly found the argument of the Gaussian to be $\left(\delta+\cos^2\frac{\theta}{2}\right)^2$.

The same calculation can be performed for an atom initially in its excited state. Looking at the stationary point of the recursion relation of Eq. \eqref{eq:recurrence relation beyond transcoherent excited}, namely,
\eq{
\cos\frac{\theta}{2}\cos\frac{\Omega_{n+1}t}{2}-\sin\frac{\theta}{2}\sin\frac{\Omega_{n}t}{2}=0,
} we now arrive at the condition
\eq{
\cos\left(\frac{\theta}{2}+\frac{\Omega_{\tilde{n}t}}{2}\right)=0\qquad \Rightarrow \qquad \Omega_{\tilde{n}}t\approx \theta+\pi.
\label{eq:stationary point excited}
} This similarly corresponds to the classical scenario in which a pulse area of $\theta+\pi$ (i.e., rotating from $\ket{\mathrm{e}}$ to $\ket{\mathrm{g}}$ to $\ket{\theta}$) is achieved when the Rabi frequency and total interaction time satisfy $\Omega_{\bar{n}}t=\theta+\pi$.

Substituting this new condition into Eq. \eqref{eq:recurrence relation beyond transcoherent excited} leads to the expansion:
\eq{
-\tan\frac{\theta}{2}\frac{\sin\frac{\Omega_{n}t}{2}}{ \cos\frac{\Omega_{n+1}t}{2}  }
\approx 1-\frac{\theta+\pi}{2\tilde{n}\sin\theta}\left(\delta+\cos^2\frac{\theta}{2}\right).
} Using this for the approximate difference equation leads to the Gaussian distribution
\eq{
|\psi_{\tilde{n}+\delta}|^2\approx |\psi_{\tilde{n}}|^2 \exp\left[
-\frac{\theta+\pi}{2\tilde{n}\sin\theta}\left(\delta+\cos^2\frac{\theta}{2}\right)^2
\right].
} This differs from Eq. \eqref{eq:Gaussian fround ground} by an innocuous-looking addition of $\pi$ that is responsible for a number of important properties (as well as the stationary point being shifted from $\tilde{n}$ by 1).

\subsection{Discussion}~
The states defined by Eqs. \eqref{eq:recurrence relation beyond transcoherent ground} and \eqref{eq:recurrence relation beyond transcoherent excited} directly generalize the transcoherent states of Ref. \cite{GoldbergSteinberg2020}. Transcoherent states, in the large-$\bar{n}$ limit, enact $\tfrac{4k+1}{2}\pi$ pulses on state $\ket{\mathrm{g}}$ because their photon-number distributions are squeezed by $\tfrac{4k+1}{2}\pi$ and $\tfrac{4k+3}{2}\pi$ pulses on state $\ket{\mathrm{e}}$ because their photon-number distributions are squeezed by $\tfrac{4k+3}{2}\pi$.
We can now understand from where these factors truly arise: number squeezing by $\sinc \theta$ leads to pulse areas of $\theta$ on states $\ket{\mathrm{g}}$ and by $\left(\theta+\pi\right)^{-1}\sin\theta=-\sinc\left(\theta+\pi\right)$ leads to pulse areas of $\theta+\pi$ on states $\ket{\mathrm{e}}$. The properties of the $\sinc$ function are responsible for the forms of the viable solutions to optimally delivering arbitrary pulse areas.

Variances cannot be negative. The $\sinc$ function, however, flips its sign periodically with period $\pi$. This means that the only pulses that can be delivered to state $\ket{\mathrm{g}}$ are those with
\eq{
(2k)\pi \leq \theta\leq (2k+1)\pi\quad \Leftarrow\quad \sigma^2=\bar{n}\sinc\theta
\label{eq:theta constraints from ground}
} and to $\ket{\mathrm{e}}$ are those with
\eq{
(2k+1)\pi \leq \theta+\pi\leq (2k+2)\pi\quad \Leftarrow\quad \sigma^2=-\bar{n}\sinc\left(\theta+\pi\right),
\label{eq:theta constraints from excited}
} where $k\in\mathds{N}_0$. These ranges, together with Eqs. \eqref{eq:stationary point ground} and \eqref{eq:stationary point excited}, cover all of the classically allowed possibilities for pulse areas, now in a fully quantized regime. The periodic maxima of these functions correspond to the transcoherent states, the negative regions explain why certain pulse areas are only accessible to atoms initially in their ground or excited states, and the property $\left|\sinc \theta\right|\leq 1$ implies that only photon-number squeezing, not photon-number broadening, is useful for gaining quantum advantages in generating atomic states with arbitrary coherence properties.
This also explains why we chose to retain the $-$ sign from Eq. \eqref{eq:recurrence relation beyond transcoherent excited} in Eq. \eqref{eq:stationary point excited} and similarly how we chose the signs of the terms in Eq. \eqref{eq:stationary point ground}:  the solutions found from the alternate choices of signs lead to minima in the recursion relations instead of maxima, corresponding to regions where the $\sinc$ functions are negative, which do not lead to valid solutions.

We can inspect a number of limits to ensure that these states behave sensibly. In the limit of a pulse area of $0$, where we desire no change in the state of the system, the best states have no number squeezing: $\sigma^2=\bar{n}$. Moreover, these states require an interaction time satisfying $\Omega_{\tilde{n}}t=0$, so $\bar{n}=0$, and the optimal solution is that the field is in its vacuum state, which is the trivial case of a coherent state with no photon-number squeezing. This solution works for atoms initially in either state $\ket{\mathrm{g}}$ or $\ket{\mathrm{e}}$; for the latter, the pulse area has $\theta+\pi=0$, which seems like it imparts a negative photon-number variance because $-\sinc 0=-1$, but this is not a problem because the product of this negative squeezing factor and $\bar{n}=0$ still vanishes.

To achieve a pulse area $l\pi$ for $l\in\mathds{N}$, a variance of zero is required, corresponding to the zeroes of the $\sinc$ function. Equivalently, the only field states that exactly generate $\pi$ pulses, $2\pi$ pulse\aaron{s}, and so on are those that are ``infinitely'' photon-number-squeezed coherent states: number states. This directly accords with the eigenstates of the JCM Hamiltonian $\ket{\pm,n}$ found in Eq. \eqref{eq:JCM eigenstates}: when the joint system begins in either $\ket{n}\otimes\ket{\mathrm{e}}$ or $\ket{n+1}\otimes\ket{\mathrm{g}}$, an interaction time of $\Omega_n t=(2l+1)\pi$ swaps the excitation between the field and the atom, while an interaction time of $\Omega_n t=(2l)\pi$ returns the excitations to their original starting points.

Complementing the properties of the $\sinc$ function, there is another reason why the ideal field states only exist for the pulse areas described by Eqs. \eqref{eq:theta constraints from ground} and \eqref{eq:theta constraints from excited}. An ideal pulse acting on $\ket{\mathrm{g}}$ must undergo a $(2k)\pi$ pulse in the lowest-excitation manifold and $(2k+1)\pi$ pulse in the highest. The pulse area for the \textit{average} photon number $\bar{n}$ must therefore always be between $(2k)\pi$ and $(2k+1)\pi$, never between $(2k+1)\pi$ and $(2k+2)\pi$; the converse holds for atoms initially in state $\ket{\mathrm{e}}$. This is why, for example, a perfect $\tfrac{\pi}{2}$ pulse can never be applied to state $\ket{\mathrm{e}}$, which must instead experience a perfect $\tfrac{3}{2}\pi$ pulse. Controlling the pulse areas of the lowest- and highest-excitation sections is paramount for ideal coherence transfer.

The idea of tailoring pulses such that the highest-excitation manifold undergoes a $\pi$ pulse was recently explored in a different context Ref. \cite{Liuetal2021constructing}.
There, this paradigm was used to create a universal set of quantum operations that could be used for quantum computation. We thus stress the importance of using the fully quantized JCM to surmount information leakage in light-matter-interaction protocols.

A useful property of transcoherent states and beyond is that the field states experience less backaction from the interaction with the atom than standard coherent states. This due to there being less residual atom-field entanglement after the interaction with the ideal states, so that the latter remain highly pure. Inspecting the evolution of $\expct{\ket{m}\bra{n}}$ using Eqs. \eqref{eq:JCM from ground} and \eqref{eq:recurrence relation beyond transcoherent ground}, for example, we find
\eq{
\psi_n^*\psi_m\to \psi_n^*\psi_m\frac{\cos\frac{\Omega_{n-1}t}{2}\cos\frac{\Omega_{m-1}t}{2}}{\cos^2\frac{\theta}{2}},
} which remains mostly unchanged for $m\sim n\sim\bar{n}$ where $\Omega_{\bar{n}}t\approx\theta$.
The reduced backaction for transcoherent states and beyond allows them to be repeatedly used as ``catalysts'' for transferring coherence to the atoms before eventually running out of energy and coherence to impart and degrading through repeated interactions \cite{GoldbergSteinberg2020}.\footnote{C.f. quantum catalysis as studied in the JCM in Ref. \cite{Messingeretal2020}.} Therefore, the cost associated with producing a transcoherent state should, in some sense, be reduced by a factor of the number of times that state can be used for practical coherence transfer.

\begin{figure}
    \centering
    \includegraphics[width=0.65\columnwidth]{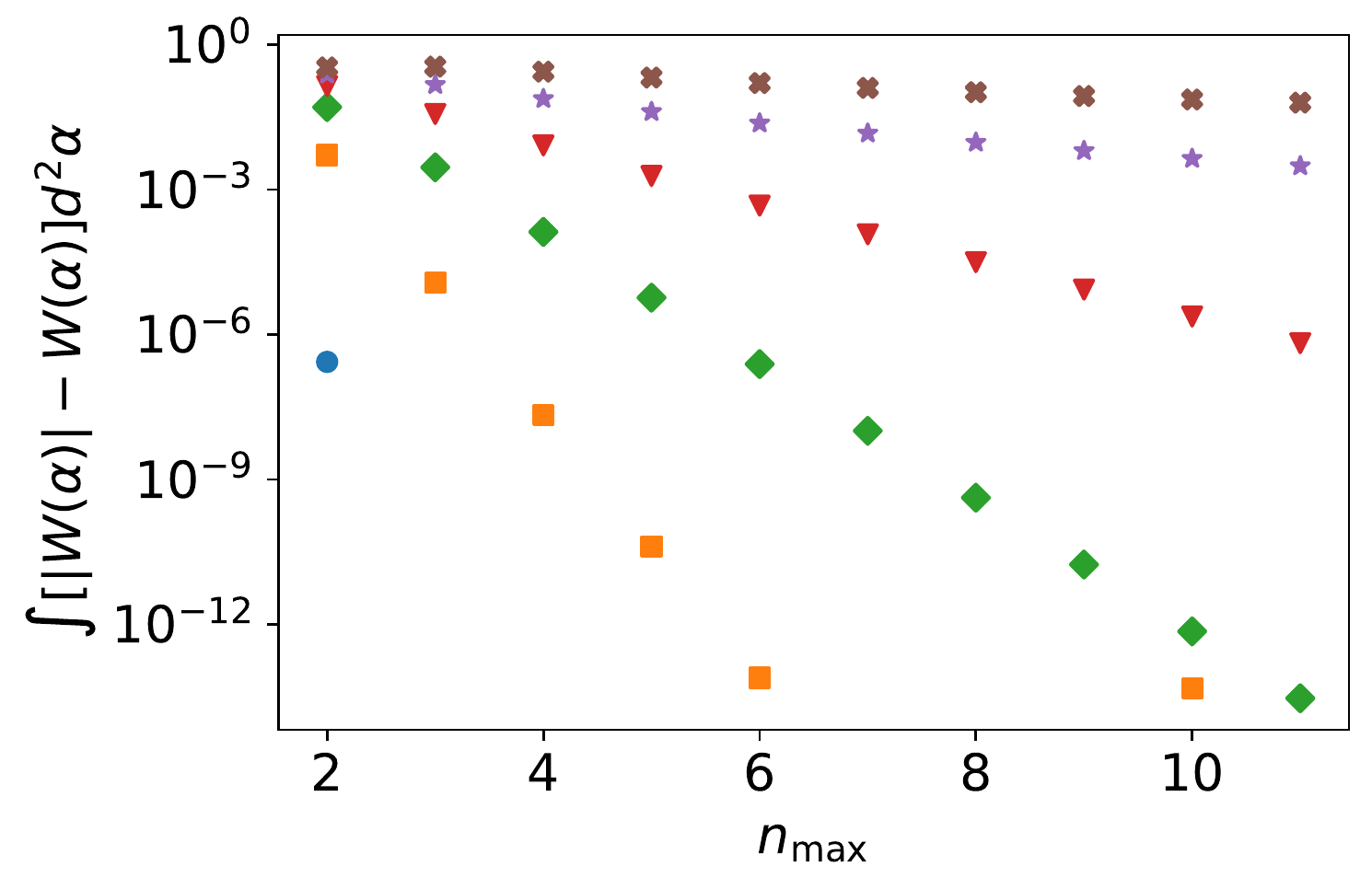}
    \caption{\aaron{Negativities of the Wigner functions for the ideal states. Plotted are the values computed for small values of $n_{\mathrm{max}}$, which are already seen to decay rapidly, indicating that the states are highly Gaussian and increase rapidly in Gaussianity. The same pulse areas are considered as in Fig. \ref{fig:beyond transcoherent from g}: blue dots for $\pi/8$, orange squares for $2\pi/8$, green diamonds for $3\pi/8$, red triangles for $4\pi/8$, purple stars for $5\pi/8$, and brown Xs for $6\pi/8$. The numerical integration procedure has a difficult time computing the negativities when they are tiny, as evidenced by the $\pi/8$ and $2\pi/8$ pulses with larger $n_{\mathrm{max}}$.}}
    \label{fig:Wigner negativities}
\end{figure}

\aaron{How easy is it to generate such ideal states? In the limit of large numbers of photons, photon-number-squeezed states are readily approximated by so-called Gaussian states, which are routinely prepared in the laboratory. Gaussian states are those whose Wigner distribution $W(\alpha)$ is nonnegative for all phase-space coordinates $\alpha$ and comprise displaced and squeezed thermal states for a single mode; since our states our pure, if they have positive Wigner functions then they must be squeezed displaced vacuum states. We plot in Fig. \ref{fig:Wigner negativities} the integrated negativity of the Wigner function \cite{KenfackZyczkowski2004} for a variety of rotation angles $\theta$ and the smallest 10 values of $n_{\mathrm{max}}$. The negativities are already seen to decay exponentially with $n_{\mathrm{max}}\sim \bar{n}\left(\tfrac{\pi}{\theta}\right)^2$, decaying most rapidly for smallest $\theta$, demonstrating how well Gaussian states can approximate the ideal states and how amenable such state generation can be. We calculated in Ref. \cite{GoldbergSteinberg2020} that only 2 dB of quadrature squeezing is required for our idealized $\tfrac{\pi}{2}$, which is routinely achieved \cite{Wuetal1986}. This now increases to $-10\log_{10}(\sinc\theta)$ for $\theta$ pulses, which can be seen because squeezed states with large mean photon number have photon-number variances $\bar{n} 10^{-S/10}$ for squeezing $S$ expressed in dB \cite{LoudonKnight1987}, meaning that 10 dB of squeezing could achieve the ideal pulses for angles up until $\theta\approx2.85$ and 15 dB (achieved by Ref. \cite{Vahlbruchetal2016}) up to $\theta\approx 3.04$. For the small values of $n_{\mathrm{max}}$ and $\bar{n}$ for which the ideal states are not as well approximated by Gaussian states, they can more readily be prepared using methods from circuit quantum electrodynamics, where certain states with particular photon-number distributions have been created beyond $\bar{n}=3$ \cite{Raimondetal2001}.}
% \begin{figure}
%     \centering
%     \includegraphics[width=0.65\columnwidth]{Negativities_vs_energy_changing_theta.eps}
%     \caption{\aaron{Negativities of the Wigner functions for the ideal states. Plotted are the values computed for the smallest values of $\bar{n}$, which are already seen to decay rapidly, indicating that the states are highly Gaussian and increase rapidly in Gaussianity. The same pulse areas are considered as in Fig. \ref{fig:beyond transcoherent from g}: blue dots for $\pi/8$, orange squares for $2\pi/8$, green diamonds for $3\pi/8$, red triangles for $4\pi/8$, purple stars for $5\pi/8$, and brown Xs for $6\pi/8$. The numerical integration procedure has a difficult time computing the negativities when they are tiny, as evidenced by the $\pi/8$ and $2\pi/8$ pulses with larger $n_{\mathrm{max}}$.}}
%     \label{fig:Wigner negativities energy}
% \end{figure}

\section{Optimal field states for generating $\Theta$ pulses on arbitrary atomic states}~
Transcoherent states and their generalizations in Section \ref{sec:perfectly generating arbitrary coherence} are the unique optimal field states that generate arbitrary atomic states $\ket{\theta}$ in arbitrarily short times from atoms initially in state $\ket{\mathrm{g}}$ or $\ket{\mathrm{e}}$. For quantum information protocols, one often seeks a transformation that transforms arbitrary initial atomic states in the same way. The transcoherent states offer a method for enacting the ideal rotation by $\tfrac{\pi}{2}$ about the $y$ axis on the Bloch sphere:
\eq{
\begin{pmatrix}
\ket{\mathrm{g}}\\\ket{\mathrm{e}}
\end{pmatrix}\to\frac{1}{\sqrt{2}}\begin{pmatrix}
1&1\\-1&1
\end{pmatrix}\begin{pmatrix}
\ket{\mathrm{g}}\\\ket{\mathrm{e}}
\end{pmatrix},
\label{eq:general pi/2 pulse}
} equivalent to a $\tfrac{\pi}{2}$ pulse, on either of the initial states $\ket{\mathrm{g}}$ or $\ket{\mathrm{e}}$. This is similar to the Hadamard transformation, which is also useful for generating coherence.
What is the optimal field state for enacting a $\tfrac{\pi}{2}$ transformation on arbitrary initial states $\ket{\theta,\phi}$ to the transformed states
\eq{
\ket{\theta,\phi}_{\frac{\pi}{2}}\equiv
\frac{\cos\frac{\theta}{2}-\eu^{\iu\phi}\sin\frac{\theta}{2}}{\sqrt{2}}\ket{\mathrm{g}}+\frac{\cos\frac{\theta}{2}+\eu^{\iu\phi}\sin\frac{\theta}{2}}{\sqrt{2}}\ket{\mathrm{e}}?
\label{eq:theta phi state after Hadamard}
} It is impossible to do this perfectly, in contrast to transcoherent states and their generalizations in Sec. \ref{sec:perfectly generating arbitrary coherence} that can do this perfectly, because the highest-excitation manifold must undergo a $(2k)\pi$ pulse for most initial states $\ket{\theta,\phi}$ but a $(2k+1)\pi$ pulse for initial state $\ket{\mathrm{e}}$. We thus seek states that perform the best on average.

A straightforward way of determining the success of creating the state depicted in Eq. \eqref{eq:theta phi state after Hadamard} is as follows:
we begin with some state $\ket{\theta,\phi}$ and some initial field state, evolve the joint system using the JCM, measure the overlap of the evolved state $\ket{\Psi(t)}$ with $\ket{\theta,\phi}_{\frac{\pi}{2}}$, and average the result over all initial atomic state angles $\theta$ and $\phi$. The result should depend on the initial field state, so we can ask what field state maximizes the resulting averaged fidelity, equivalent to the averaged success probability. 

By combining Eqs. \eqref{eq:JCM from ground} and \eqref{eq:JCM from excited}, we learn that a state $\sum_n \psi_n\ket{n}\otimes\ket{\theta,\phi}$ evolves to
\eq{
\ket{\Psi(t)}=\sum_{n=0}^\infty\ket{n}\otimes\left(\cos\frac{\theta}{2}\ket{G_n}+\sin\frac{\theta}{2}\eu^{\iu\phi}\ket{E_n}\right),
\label{eq:transformed joint state from arbitrary initial}
} where we have defined the atomic states
\eq{
\ket{G_n}&=\psi_n\cos\frac{\Omega_{n-1}t}{2}\ket{\mathrm{g}}-\iu \psi_{n+1}\sin\frac{\Omega_n t}{2}\ket{\mathrm{e}} ,\\
\ket{E_n}&=\psi_n\cos\frac{\Omega_n t}{2}\ket{\mathrm{e}}
-\iu \psi_{n-1}\sin\frac{\Omega_{n-1} t}{2}\ket{\mathrm{g}}.
}
Comparing this result to the desired state $\ket{\theta,\phi}_{\frac{\pi}{2}}$, we observe that a strongly peaked photon-number distribution around $\bar{n}$ satisfying \eq{
\psi_{\bar{n}}\cos\frac{\Omega_{\bar{n}-1}t}{2}\approx \psi_{\bar{n}}\cos\frac{\Omega_{\bar{n}}t}{2}
&
\approx \iu \psi_{{\bar{n}}-1}\sin\frac{\Omega_{\bar{n}-1}t}{2}%\\&
\approx
-\iu \psi_{{\bar{n}}+1}\sin\frac{\Omega_{\bar{n}}t}{2}
\label{eq:desired distribution properties}
} would be highly beneficial for performing the transformation of Eq. \eqref{eq:general pi/2 pulse}. This seems to suggest an optimal interaction time conforming to the classical relationship $\Omega_{\bar{n}}t=\tfrac{\pi}{2}$.

There are a few considerations to calculating the averaged fidelity. To be viable for arbitrary initial states, all angles of the Bloch sphere should be equally weighted in the average.
If the azimuthal angle of the initial atomic state were known, on the other hand, one could integrate only over the polar coordinate, after waiting an appropriate free evolution such that $\phi\to 0$. 
It is not obvious that there will be a single unique solution: some points on the Bloch sphere are hardly rotated by a $\tfrac{\pi}{2}$ pulse because they are close to the $\pm y$-axis thereof.
Indeed, these processes yield different ideal states and will be considered in turn.

\subsection{Averaging fidelity over all initial atomic states}~

We calculate the squared overlap averaged over the entire surface of the Bloch sphere for $\tfrac{\pi}{2}$ pulses in Appendix \ref{app:averaging fidelity} to find, for an optimal phase relationship $\psi_n \psi_{n+1}^*=-\iu\left|\psi_n \psi_{n+1}\right|$,
\eq{
\mathcal{F}=&\frac{1}{2}+\frac{1}{6}\sum_n\left|\psi_n\right|^2 \cos\frac{\Omega_n t}{2}\cos\frac{\Omega_{n-1}t}{2}%\\
%&
+\left|
\psi_{n-1}\psi_{n+1}\right|\sin\frac{\Omega_n t}{2}\sin\frac{\Omega_{n-1}t}{2}
\\
&+ \left|\psi_n \psi_{n+1}\right|\sin\frac{\Omega_{n}t}{2}\left(
2\cos\frac{\Omega_n t}{2}+
\cos\frac{\Omega_{n-1}t}{2}+
\cos\frac{\Omega_{n+1}t}{2}
\right).
\label{eq:averaged fidelity general}
} What photon-number distributions and what times optimize this averaged fidelity?

It is clear from counting the terms in Eq. \eqref{eq:averaged fidelity general} that achieving distributions resembling Eq. \eqref{eq:desired distribution properties} would allow for averaged fidelities approaching unity. But the relationships in Eq. \eqref{eq:desired distribution properties} compete with each other: a narrow photon-number distribution ensures that the Rabi frequencies coincide \aaron{to ensure the photon-number distribution's probability to be concentrated there}, while a broad distribution increases the overlap between adjacent photon-number coefficients. We are thus faced with the same problem of optimizing the width of the photon-number distribution that we faced in finding the transcoherent states.

\begin{figure}
    \centering
    \includegraphics[width=0.65\columnwidth]{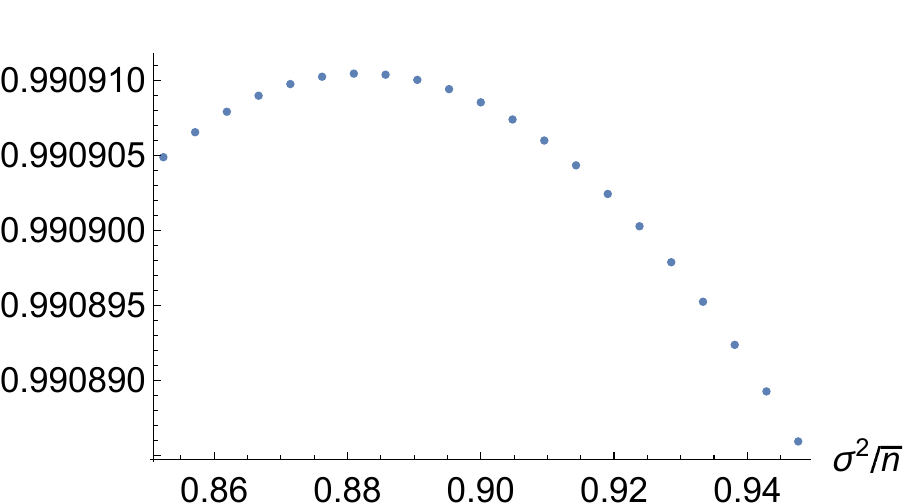}
    \caption{Average fidelity $\mathcal{F}$ for rotating atoms in arbitrary initial states by $\tfrac{\pi}{2}$ calculated from Eq. \eqref{eq:averaged fidelity general} using states of the form of Eq. \eqref{eq:Gaussian photon number distribution} with various variances. The optimal photon-number variances are scaled by that of coherent light with $\sigma^2=\bar{n}$. The cutoff point was chosen to be $n_{\mathrm{max}}=400$; here, $\bar{n}=20$.}
    \label{fig:nbar20}
\end{figure}

Writing the photon-number distribution as
\eq{
|\psi_n|^2\propto\exp\left[-\frac{\left(n-\bar{n}\right)^2}{2\sigma^2}\right]
\label{eq:Gaussian photon number distribution}
} up until some manual cutoff $n_{\mathrm{max}}\gg \bar{n}+\sigma$, we can optimize over the variances $\sigma^2$ for various values of the average photon number $\bar{n}$. Intriguingly, we find that, for sufficiently large $\bar{n}$, the optimal variance is always \textit{slightly} number squeezed, approaching
\eq{
\sigma_{\mathrm{optimal}}^2\approx  0.9\bar{n}.
\label{eq:optimal variance for average pi over 2}
} This is depicted in Fig. \ref{fig:nbar20}.%, \ref{fig:nbar80}, and \ref{fig:nbar320}.
% \begin{figure}
%     \centering
%     \includegraphics[width=0.65\columnwidth]{nmax_400_nbar_80_ave_fids}
%     \caption{Same as Fig. \ref{fig:nbar20} but with $\bar{n}=80$.}
%     \label{fig:nbar80}
% \end{figure}
% \begin{figure}
%     \centering
%     \includegraphics[width=0.65\columnwidth]{nmax_400_nbar_320_ave_fids}
%     \caption{Same as Fig. \ref{fig:nbar20} but with $\bar{n}=320$.}
%     \label{fig:nbar320}
% \end{figure}

% An ideal field state would satisfy the relation
% \eq{
% \frac{\cos\frac{\theta}{2}-\eu^{\iu\phi}\sin\frac{\theta}{2}}
% {\cos\frac{\theta}{2}+\eu^{\iu\phi}\sin\frac{\theta}{2}}=
% \frac
% {\psi_n \cos\frac{\Omega_{n-1}t}{2}\cos\frac{\theta}{2}-\iu \psi_{n-1} \sin\frac{\Omega_{n-1}t}{2}\sin\frac{\theta}{2}\eu^{\iu\phi}}
% {-\iu \psi_{n+1} \sin\frac{\Omega_{n}t}{2}\cos\frac{\theta}{2}+ \psi_{n} \cos\frac{\Omega_{n}t}{2}\sin\frac{\theta}{2}\eu^{\iu\phi}}.
% } Given that $\psi_0$ is determined by normalization and $\psi_{-1}$ does not exist, this defines a recurrence relation for the coefficients $\psi_n$ up until some manual cutoff point $n_{\mathrm{max}}$. We can choose some time points $\Omega_0 t$, calculate the coefficients from this recurrence relation, and inspect the properties of the resulting state for various choices of $\theta$. Perhaps they may agree? In fact, they seem to prefer different amounts of squeezing depending on the initial choices of the angles, with perhaps a $\sin\theta-\cos\theta$ dependence.

\begin{figure}
    \centering
    \includegraphics[width=0.65\columnwidth]{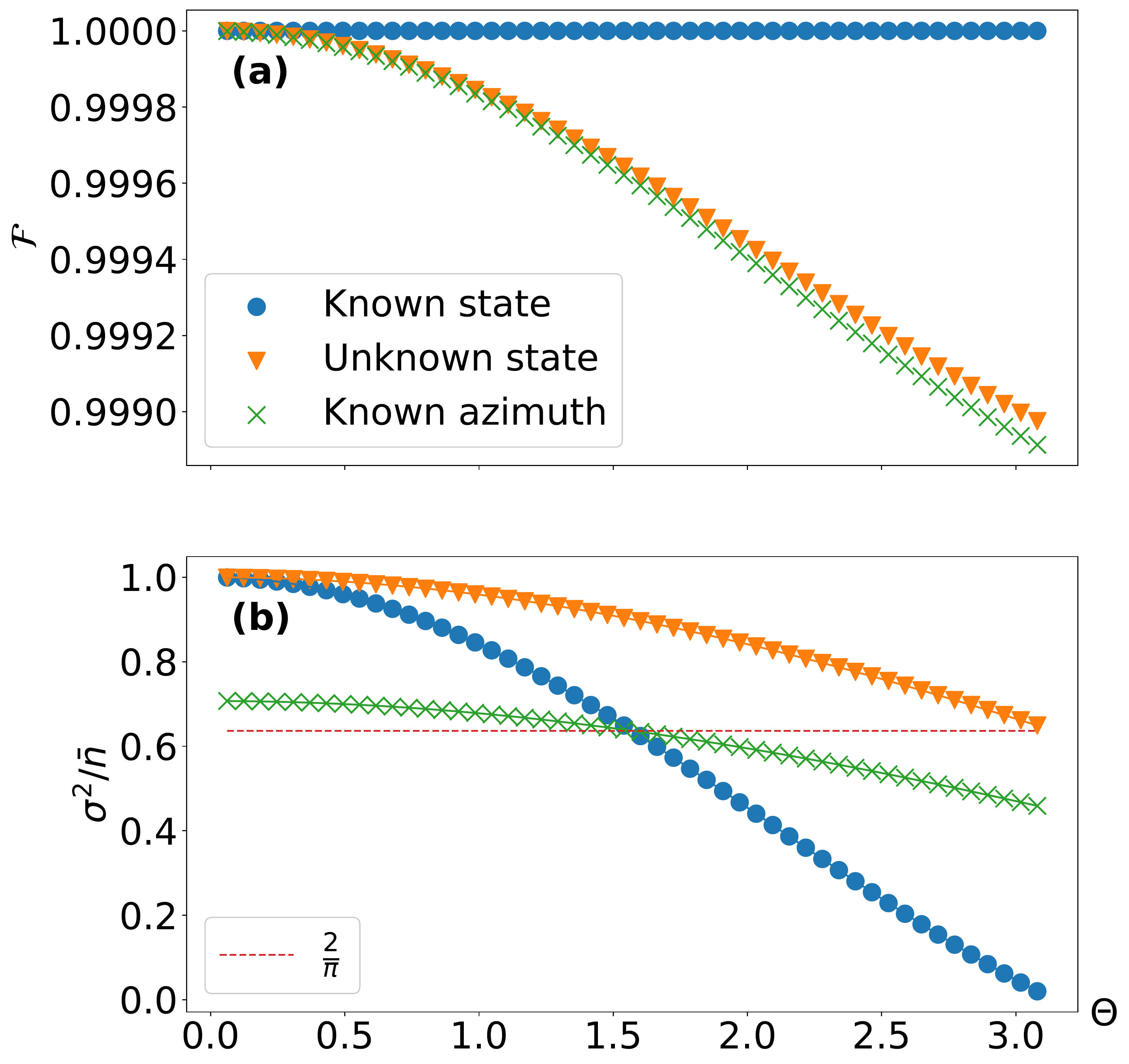}
    \caption{Optimal fidelity and squeezing for field states imparting $\Theta$ pulses on atoms with known (blue dots), unknown (orange triangles), and partially known (green $\times$s) initial states. Each point represents a different value of $\Theta$ for which the average fidelities were optimized over all possible variances and a fixed energy $\bar{n}=500$. \textbf{(a)} The average fidelities achieved are all large because we have used field states with large intensities. Known-state atoms can be always be perfectly transformed, while it is easier to achieve shorter pulses with smaller $\Theta$ for atoms in unknown states. \textbf{(b)} The optimal variances are plotted in units of $\bar{n}$. Plotted on top are the curves $\sinc\Theta$ (blue), $\sinc\tfrac{\Theta}{2}$ (orange), and $\sqrt{1/2}\sinc\tfrac{\Theta}{2}$ (green). The blue and green curves intersect at $\Theta=$$\tfrac{\pi}{2}$ when $\sigma^2=\tfrac{2\bar{n}}{\pi}$ (red, dashed).}
    \label{fig:optimal pulses all situations}
\end{figure}

We can repeat this calculation to find the optimal squeezed \aaron{state} to achieve any $\Theta$ pulse when averaged over all initial atomic states. The calculation yields a more cumbersome version of Eq. \eqref{eq:averaged fidelity general} and is given in Appendix \ref{app:averaging fidelity any pulse}. Optimizing this expression numerically over Gaussian field states, we find the intriguing relationship (Fig. \ref{fig:optimal pulses all situations})
\eq{
\sigma^2_{\mathrm{optimal}}= \bar{n}\sinc\frac{\Theta}{2},
\label{eq:sinc variance any atom}
} which explains the $0.9$ \aaron{factor} found in Eq. \eqref{eq:optimal variance for average pi over 2}. The maximum achievable fidelity decreases with $\Theta$ for a given $\bar{n}$ and is notably less than the perfect fidelities achievable when the atom is initially in its ground state. In fact, comparing this expression with Eq. \eqref{eq:sinc variance}, we see that the optimal field state for delivering a $\Theta$ pulse to an unknown atomic state has the same amount of squeezing as the optimal field state for delivering a $\tfrac{\Theta}{2}$ pulse to an atom in its ground state. While we only speculate on the origin of this conclusion, we are confident that it arises from some averaging of the distance that an atom must traverse on the Bloch sphere, which ranges from $0$ to $\Theta$; i.e., the average atom must rotate half as far as a ground-state atom during a $\Theta$ pulse, so the average atom requires a $\tfrac{\Theta}{2}$ pulse.

% \begin{figure}
%     \centering
%     \includegraphics[width=0.65\columnwidth]{fidelity_and_variance_all_phi_nbar500.eps}
%     \caption{Optimal average fidelity and squeezing for field states imparting $\Theta$ pulses on atoms with completely unknown states. Each dot represents a different value of $\Theta$ for which the average fidelities were optimized over all possible variances and a given, fixed $\bar{n}=500$. \textbf{(a)} The average fidelities achieved are all large because we have used field states with large intensities. Shorter pulses with smaller $\Theta$ are easier to achieve than longer pulses \textbf{(b)} The optimal variances are plotted in units of $\bar{n}$. Plotted on top of the dots is the curve $\sigma^2=\bar{n}\sinc\tfrac{\Theta}{2}$.}
%     \label{fig:optimal pulses all phi}
% \end{figure}

\subsection{Averaging fidelity over initial states with known azimuth}~
We next take the case where the initial azimuthal angle of the atom is known; this is equivalent to taking $\phi=0$ instead of averaging over that coordinate. As calculated in Appendix \ref{app:averaging fidelity any pulse fixed phi}, the fidelity for a $\tfrac{\pi}{2}$ pulse, averaged over all initial values of the atom's polar coordinate $\theta$, is
\eq{
\mathcal{F}=&\frac{1}{2}
+\frac{1}{4}\sum_n \left|\psi_n \psi_{n+1}\right|\sin\frac{\Omega_n t}{2}
\left(2\cos\frac{\Omega_n t}{2}+
\cos\frac{\Omega_{n-1} t}{2}+
\cos\frac{\Omega_{n+1} t}{2}\right)
.
} 
By numerically maximizing this quantity over all states with Gaussian photon-number distributions, we find that the optimal field state is a coherent state that is number squeezed by $\tfrac{\pi}{2}$. This is exactly the same result as for transcoherent states, even though the corresponding quantity in Ref. \cite{GoldbergSteinberg2020} [Eq. (13) there] has a single cosine term instead of all three, so we proceed with the same method to justify our optimal solution. To maximize the fidelity, we need to maximize the inner product between vectors with components $\psi_{n+1} \sin\tfrac{\Omega_n t}{2}$ and $\psi_{n}\left(2\cos\frac{\Omega_n t}{2}+
\cos\frac{\Omega_{n-1} t}{2}+
\cos\frac{\Omega_{n+1} t}{2}\right)/4$. By the Cauchy-Schwarz inequality, this is achieved when the vectors are parallel, satisfying
\eq{
\psi_{n+1} \sin\tfrac{\Omega_n t}{2}=\frac{\psi_{n}}{4}\left(2\cos\frac{\Omega_n t}{2}+
\cos\frac{\Omega_{n-1} t}{2}+
\cos\frac{\Omega_{n+1} t}{2}\right).
} This generates a recursion relation for the ideal state coefficients that can be expanded about their peak at $\bar{n}$ for an evolution time $\Omega_{\bar{n}}t=\tfrac{\pi}{2}$:
\eq{
\frac{2\cos\frac{\Omega_{\bar{n}+\delta} t}{2}+
\cos\frac{\Omega_{\bar{n}+\delta-1} t}{2}+
\cos\frac{\Omega_{\bar{n}+\delta+1} t}{2}}{4\sin\tfrac{\Omega_{\bar{n}+\delta} t}{2}}\approx 1-\frac{\pi\delta}{4\bar{n}}.
} %OR
%$\Omega_{0}t\sqrt{\bar{n}}=\tfrac{\pi}{2}$:
%\eq{
%\frac{2\cos\frac{\Omega_{\bar{n}-1+\delta} t}{2}+
%\cos\frac{\Omega_{\bar{n}-1+\delta-1} t}{2}+
%\cos\frac{\Omega_{\bar{n}-1+\delta+1} t}{2}}{4\sin\tfrac{\Omega_{\bar{n}-1+\delta} t}{2}}\approx 1-\frac{\pi\delta}{4\bar{n}}.
%}
We select a probability distribution satisfying the approximate difference equation
\eq{
\frac{\psi_{\bar{n}+\delta}-\psi_{\bar{n}}}{\delta}\approx -\frac{\pi\delta}{4\bar{n}}\psi_{\bar{n}} 
,
} whose solution is the photon-number-squeezed Gaussian distribution
\eq{
\psi_{\bar{n}+\delta}\approx \psi_{\bar{n}}\exp\left(-\frac{\delta^2}{4\sigma^2}\right),\quad \sigma^2=\frac{2\bar{n}}{\pi}.
}
We thus observe that the best states for \textit{exactly} producing a $\tfrac{\pi}{2}$ pulse and for \textit{on average} producing a $\tfrac{\pi}{2}$ pulse are the same, so long as the azimuthal coordinate of the initial atomic state is known. 

% \begin{figure}
%     \centering
%     \includegraphics[width=0.65\columnwidth]{fidelity_and_variance_fixed_phi_nbar500.eps}
%     \caption{Optimal average fidelity and squeezing for field states imparting $\Theta$ pulses on atoms with unknown polar coordinate. Each dot represents a different value of $\Theta$ for which the average fidelities were optimized over all possible variances and a given, fixed $\bar{n}=500$. \textbf{(a)} The average fidelities achieved are all large because we have used field states with large intensities. Shorter pulses with smaller $\Theta$ are easier to achieve than longer pulses \textbf{(b)} The optimal variances are plotted in units of $\bar{n}$, on top of which is the curve $\sigma^2=\bar{n}\sinc\tfrac{\Theta}{2}/\sqrt{2}$. The dashed line is at $\tfrac{2}{\pi}$, which is exactly the amount of squeezing required for $\tfrac{\pi}{2}$ pulses ($\Theta=\pi/2$).}
%     \label{fig:optimal pulses fixed phi}
% \end{figure}

The same calculation can be done for arbitrary pulse areas $\Theta$. Averaging the success probability for acting on atoms with $\phi=0$ and arbitrary $\theta$ (Appendix \ref{app:averaging fidelity any pulse fixed phi}), we find that the optimal variances obey (Fig. \ref{fig:optimal pulses all situations})
\eq{
\sigma^2=\frac{2\bar{n}}{\pi}\frac{\sinc \frac{\Theta}{2}}{\sinc \frac{\pi}{4}}=\bar{n}\frac{\sinc \frac{\Theta}{2}}{\sqrt{2}}.
\label{eq:sinc variance known phi}
} As usual, larger pulse areas require more photon-number squeezing, and we find that more squeezing is required than when averaging over the entire Bloch sphere.

\subsection{Discussion}~
The optimal field state for enacting a pulse area of $\Theta$ on an atomic state depends on the atomic state. We can collect some of our results: when the atom is initially in its ground state, the optimal photon-number variance is [Eq. \eqref{eq:sinc variance}] $\bar{n}\sinc\Theta$; when the atom is initially in a state with some known $\phi$ but unknown polar angle, the optimal variance is [Eq. \eqref{eq:sinc variance known phi}] $\bar{n}\sinc\tfrac{\Theta}{2}/\sqrt{2}$; and, when the atom is initially in an unknown state, the optimal variance is [Eq. \eqref{eq:sinc variance any atom}] $\bar{n}\sinc\tfrac{\Theta}{2}$. How do all of these compare with each other?

In terms of fidelity, not knowing the initial state leads to poorer performance. Surprisingly, averaging over a known azimuth leads to slightly smaller fidelities than averaging over the entire sphere. This discrepancy arises from the different Jacobian factors when integrating over a circle versus a sphere, implying that the ratio of the performances of states initially near the equator to states initially at the poles is what controls the overall success on average.

All of the scenarios require more photon-number squeezing for larger pulse areas. When the pulse area is $\tfrac{\pi}{2}$ or greater, an atom in its ground state requires the most squeezing because it has to travel the furthest, an atom oriented along a known meridian requires less squeezing on average, and an atom oriented in an unknown direction requires the least squeezing on average. That the completely unknown orientation requires the least squeezing makes sense: on average, such atoms need to traverse an angular distance of $\tfrac{\Theta}{2}$ for a rotation about some fixed axis on the Bloch sphere. That the polar-angle-unknown orientation requires less squeezing than ground-state atoms is more surprising: this implies that it requires more ``effort,'' in terms of greater squeezing, to travel from the poles of the Bloch sphere than from any other point. The cause of this discrepancy for angles other than $\Theta=\tfrac{\pi}{2}$ remains an open question for further study.

When the pulse area is less than $\tfrac{\pi}{2}$, the variances quoted above are more squeezed for the case with known $\phi$ than for atoms initially in their ground states. While this may be an empirical phenomenon, there is a fly in the ointment: it is not clear for $\Theta<\tfrac{\pi}{2}$ what the optimal relationship \eq{\varphi\equiv \arg \psi_{n+1}-\arg \psi_n} in Appendix \ref{app:averaging fidelity any pulse fixed phi}
should be for the initial field states. However, performing a multiparameter optimization over $\varphi$ and $\sigma^2$, we always find the optimal value to have $\varphi\approx\tfrac{\pi}{2}$ and thus maintain the variance relationship of Eq. \eqref{eq:sinc variance known phi}. We must then conclude that, somehow, ground-state atoms require less squeezing than others for rotations by $\Theta<\tfrac{\pi}{2}$ about a great circle and more squeezing than others for rotations by $\Theta>\tfrac{\pi}{2}$ about the same great circle. This is an intriguing phenomenon that surely deserves further research.

% This quantity to maximize looks the same as the one for the original transcoherent states [\cite{GoldbergSteinberg2020} Eq. (13)], where the $\cos\frac{\Omega_n t}{2}$ there is replaced by the sum of three cosine terms here. In fact, we find almost the same expansion
% \eq{
% \sin\frac{\Omega_{\bar{n}} t}{2}\left(2\cos\frac{\Omega_{\bar{n}} t}{2}+
% \cos\frac{\Omega_{\bar{n}-1} t}{2}+
% \cos\frac{\Omega_{\bar{n}+1} t}{2}\right)\\
% \approx 2\left[1+\frac{\pi(4-\pi)}{256\bar{n}^2}+\mathcal{O}\left(\bar{n}^{-3}\right)\right].
% } This is best matched by coherent states that are number squeezed with 
% \eq{
% \sigma^2=\frac{256\bar{n}^2}{\pi(4-\pi)}
% } so that the large-$\bar{n}$ expansion of the fidelity drops off from its maximum as $\mathcal{O}\left(\bar{n}^{-3}\right)$??? Also, these would be highly number broadened, not quite coherent. Next, we could try averaging this term and the $\psi_{n-1}\psi_n$ term, because they should have the same value for the Gaussian/variance, but if we average the two together we find that the trig terms give 2-... instead of 2+..., and that would mean the variance should be negative to balance it out, so that's no good. It's just way too broad to use this method of expanding about the peak, perhaps, because the variance seems to go as nbar^2 instead of as nbar.

%We next unravel the averaged fidelity into latitudes defined \textit{with respect to the $y$-axis of the Bloch sphere} and find the ideal states for each latitude. These are the components that collectively make up the overall averaged fidelity.

\begin{figure}
    \centering
    \includegraphics[width=0.65\columnwidth]{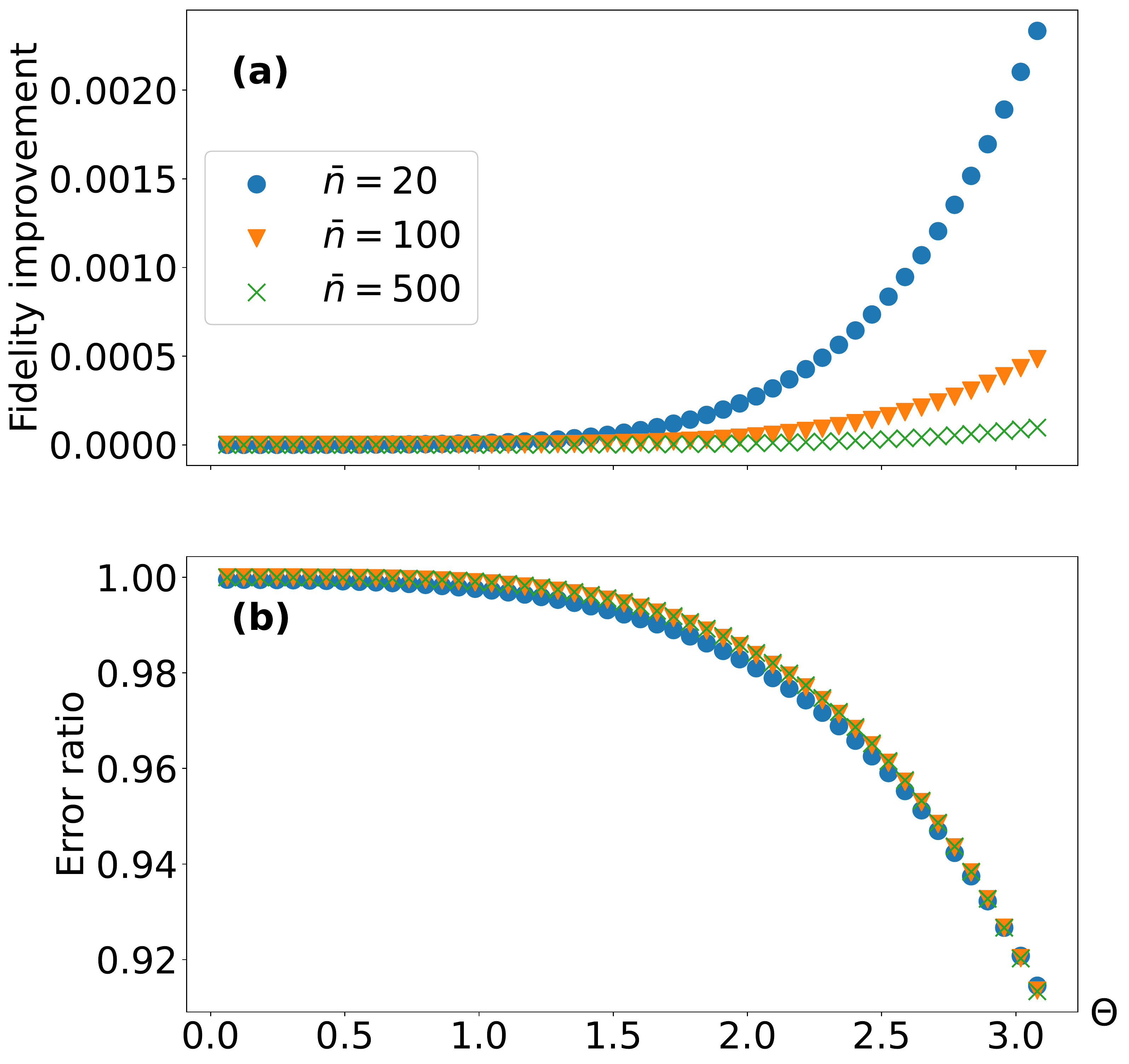}
    \caption{\textbf{(a)} Additive increase in the fidelity $\mathcal{F}$ of performing a $\Theta$ rotation on an atom averaged over all initial atomic states using light whose photon-number distribution is squeezed by an optimal amount $\sinc\tfrac{\Theta}{2}$ relative to coherent light with the same strength. The improvement is significant for larger rotation angles and smaller field-state strengths. \textbf{(b)} Multiplicative decrease in the error $1-\mathcal{F}$ of performing a $\Theta$ rotation on an atom averaged over all initial atomic states using light whose photon-number distribution is squeezed by an optimal amount $\sinc\tfrac{\Theta}{2}$ relative to coherent light with the same strength. The improvement is significant for larger rotation angles and is largely independent of the field-state strength.
    }
    \label{fig:fidelity improvement}
\end{figure}

Typically, quantum computing algorithms require the same operation to be performed on arbitrary initial states. Given the optimal field state for this purpose that is squeezed by $\sinc\tfrac{\theta}{2}$, what advantage can one acquire relative to standard quantum computing protocols that use coherent states to perform logic gates on atoms? We plot in Fig. \ref{fig:fidelity improvement} the improvement in the average fidelity [Eq. \eqref{eq:averaged fidelity general}] that one can attain using the optimally squeezed states relative to coherent states with no squeezing, comparing how this improvement changes with the energy of the field state. The fidelity improvement of squeezed states over coherent states increases quickly with the rotation angle and the improvement lessens with increasing $\bar{n}$, while the relative error decreases with rotation angle and is independent from $\bar{n}$. These imply that quantum computing applications that are limited in average photon number, that are using many $\pi$ gates, or that possess any significant error rates may benefit the most from using squeezed light to improve their logic gates.

\section{Generating $\tfrac{\pi}{2}$ pulses for collections of atoms}~
Can the transcoherent states be generalized to field states that impart optimal pulses on collections of atoms? 
A set of atoms all in the maximally coherent state $\ket{\tfrac{\pi}{2}}$ is useful for applications such as creating lasers with noise-free amplification \cite{ScullyZubairy1988}.
We investigate the collective interaction governed by the Tavis-Cummings interaction Hamiltonian \cite{TavisCummings1968}
\eq{
H_{\mathrm{TC}}=\frac{\Omega_0}{2}( \ha J_+ + \had J_-),
} where we now employ the collective excitation operators from SU(2):
\eq{
J_i=\sum_{k=1}^{2J}\mathds{I}^{(1)}\otimes\cdots\otimes\mathds{I}^{(k-1)}\otimes\sigma_i^{(k)}\otimes\mathds{I}^{(k+1)}\otimes\cdots\otimes\mathds{I}^{(2J)}.
} Here, $2J$ is the total number of atoms, where the $k$th atom has its own Pauli operators $\sigma_i^{(k)}$ and the permutation-symmetric states of the $2J$ atoms are equivalent to a single spin-$J$ particle. The transcoherent state problem begins with all of the atomic states in their collective ground state $\ket{J,-J}=\ket{\mathrm{g}}^{\otimes 2J}$ that is annihilated by $J_-$ and is an eigenstate of $J_z$ with minimal eigenvalue $-J$. A $\tfrac{\pi}{2}$ pulse acting on all atoms simultaneously would enact the transformation $\ket{\mathrm{g}}^{\otimes 2J}\to  2^{-J} (\ket{\mathrm{g}}+\ket{\mathrm{e}})^{\otimes 2J}$, where the latter is also a spin-coherent state that is an eigenstate of $J_x$ with maximal eigenvalue $J$ and can be expressed in the basis of $J_z$ eigenstates as
\eq{
\ket{J,-J}\to\frac{1}{2^J}\sum_{m=-J}^J \sqrt{\binom{2J}{m+J}}\ket{J,m}.
\label{eq:pi/2 pulse on 2J atoms}
} How can this best be performed?

\subsection{Optimal pulses for maximum coherence generation}~
We can investigate a series of field states to find which ones best impart a $\tfrac{\pi}{2}$ pulse on a collection of $2J$ atoms. Unlike in the case of the JCM, it is not convenient to write a closed-form expression for the fidelity as a function of the field-state coefficients. Instead, we choose a variety of representative field states, from which we evolve the TCM numerically using \texttt{QuTiP} \cite{Johanssonetal2012,Johanssonetal2013}. These can then be compared to the optimal final state from Eq. \eqref{eq:pi/2 pulse on 2J atoms} and optimized accordingly.

To make the optimization tractable, we choose to optimize over field states with Gaussian photon-number distributions with varying widths. This is motivated in part by the optimal states for the JCM always having Gaussian photon-number distributions, in part because Gaussian states are among the easiest to prepare experimentally, and in part because field states with sufficiently large average photon number and sufficiently localized photon-number distributions will convert $H_{\mathrm{TC}}$ into a rotation of the form
\eq{
\exp\left(-\iu H_{\mathrm{TC}}t\right)\underset{\bar{n}\gg 1}{\approx}\exp\left[-i\Omega_0 \sqrt{\bar{n}}t\left(J_x\cos\varphi+J_y\sin\varphi\right)\right],
\label{eq:strong field approx TCM rotation}
} where\aaron{,} again\aaron{,} $\varphi$ encodes the relative phases of the field-state coefficients. For a given fixed $\bar{n}$, we thus expect the fidelity to be optimized by an interaction time $\Omega_0 t\sqrt{\bar{n}}\approx $$\tfrac{\pi}{2}$.

\begin{figure}
    \centering
    \includegraphics[width=0.65\columnwidth]{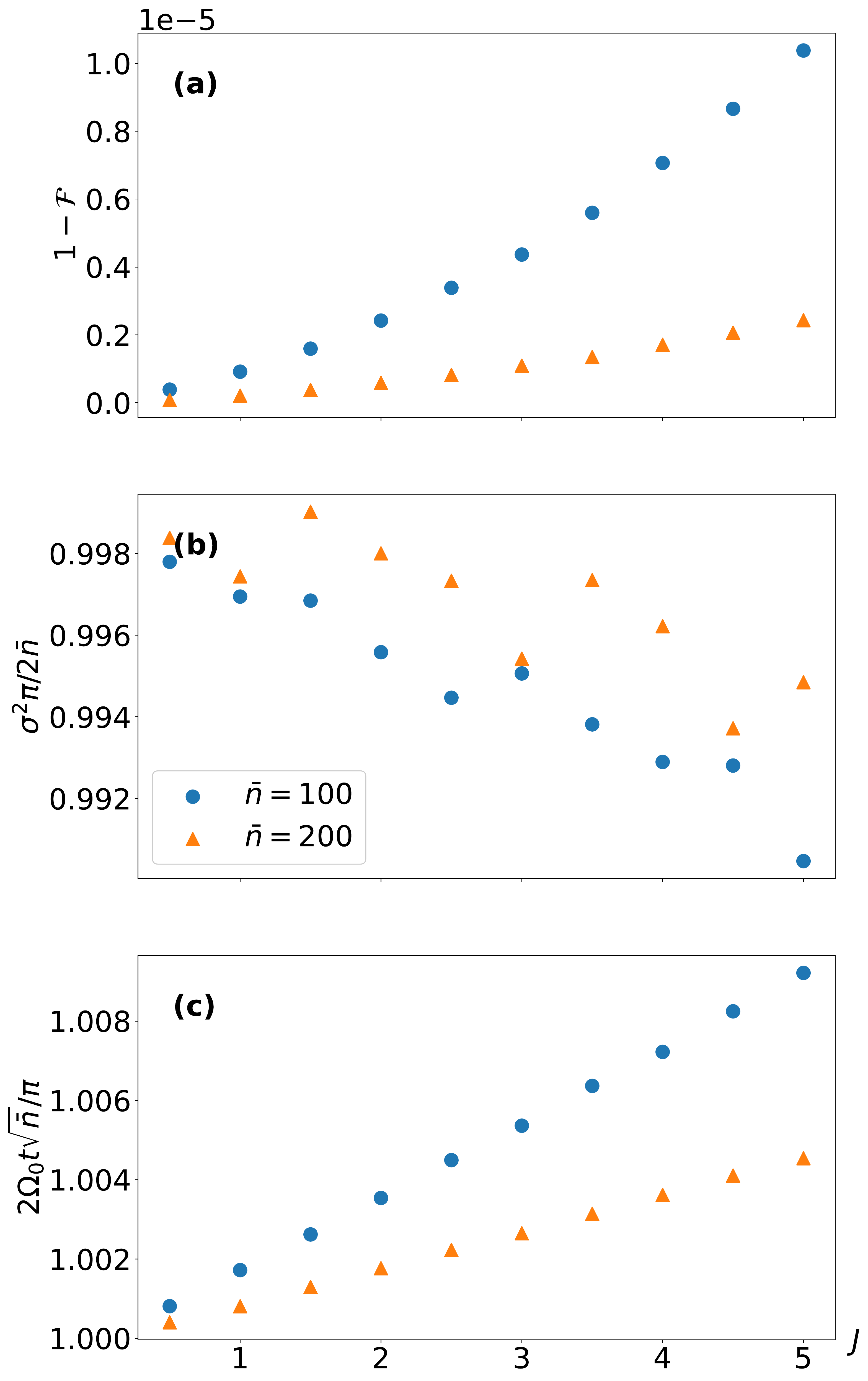}
    \caption{Optimal $\tfrac{\pi}{2}$ pulses on a collection of $2J$ atoms or another spin-$J$ system, for various average energies (blue dots and orange triangles have $\bar{n}=100$ and $\bar{n}=200$, respectively). \textbf{(a)} Error probability, i.e., unity minus the fidelity, of enacting a perfect $\tfrac{\pi}{2}$ pulse to achieve the transformation of Eq. \eqref{eq:pi/2 pulse on 2J atoms}. All of the fidelities are excellent, with larger $\bar{n}$ being more favourable and larger $J$ being less favourable. \textbf{(b)} Optimal variances for the initial field states. These all have their photon-number distributions squeezed by approximately $\tfrac{\pi}{2}$ relative to coherent states, with increased squeezing required for increasing $J$. The scatter in the plot implies that not all of the results have converged to their optimal values, which may only be achievable with larger $\bar{n}$ and longer optimization times. \textbf{(c)} Optimal interaction times to achieve the desired transformation. These are all approximately the classical times for a $\tfrac{\pi}{2}$ pulse, with a slight increase in optimal time with increasing $J$. Of note, the products of the optimal variances and times are approximately unity in these units, implying that $\sigma^2\Omega_0 t/\sqrt{\bar{n}}\approx 1$. 
    }
    \label{fig:TCM optimals}
\end{figure}

Figure \ref{fig:TCM optimals} plots the optimal field-state variances and interaction times for achieving $\tfrac{\pi}{2}$ pulses for various values of $J$ and $\bar{n}$. These parameters are the best ones found using the Nelder–Mead method implemented in \texttt{SciPy} with a variety of random seeds. As expected, the overall fidelities increase with increasing $\bar{n}$. It is perhaps unsurprising that they decrease with increasing $J$, as $J=1/2$ is the only situation in which perfect $\tfrac{\pi}{2}$ pulses can be implemented, and that the increase is quadratic in $J$ from this minimum. The optimal squeezing and time parameters follow opposite trends such that the optimal photon-number variance and interaction time obey the following relationship for a given average photon number: \eq{
\sigma_{\mathrm{optimal}}^2&\approx \frac{2\bar{n}}{\pi} \qquad \mathrm{and}\qquad
\Omega_0 t_{\mathrm{optimal}}\sqrt{\bar{n}}\approx \frac{\pi}{2} \\
&\Rightarrow\quad
\sigma_{\mathrm{optimal}}^2\Omega_0 t_{\mathrm{optimal}}\approx \sqrt{\bar{n}}.
\label{eq:optimal params TCM}
} There is also some residual dependence on $J$ that may warrant the replacement $\bar{n}\to\bar{n}-J/2+1/2$ in Eq. \eqref{eq:optimal params TCM}. \aaron{While the fidelities are seen in Fig. \ref{fig:TCM optimals} to worsen with increasing $J$, they improve with increasing $\bar{n}$, so maintaining $\bar{n}\gg J/2$ allows for highly efficient pulses for arbitrarily large $J$. This can also be seen because the field state needs to have sufficient energy to excite the atoms; when the field state's energy is highly depleted, it loses its ideal character and experiences more backaction from the interactions, similar to the decrease in its capacity for quantum catalysis \cite{GoldbergSteinberg2020}.}

\subsection{Semiclassical limit}~
The semiclassical limit of the TCM with a highly energetic field has been studied in Refs. \cite{DrobnyJex1993,Chumakovetal1994,KlimovChumakov1995,Retamaletal1997}. Those showed that, like in the JCM \cite{GeaBanacloche1990,PhoenixKnight1991a,PhoenixKnight1991b,GeaBanacloche1991,GeaBanacloche1992,GeaBanacloche2002,vanEnkKimble2002,SilberfarbDeutsch}, one cannot simply employ the replacement $\ha\to\alpha$ in $H_{\mathrm{TC}}$ in the strong-field limit as per Eq. \eqref{eq:strong field approx TCM rotation}, as this neglects possible atom-field entanglement for any finite $\bar{n}$ and wrongly predicts the final atomic state to be pure. In Ref. \cite{Retamaletal1997}, for example, we find that the TCM Hamiltonian can be approximated in the presence of a strong field with the appropriate relative phase relationships as
\eq{
\tilde{H}_{\mathrm{TC}}=-\Omega_0\sqrt{\had\ha-J+1/2}{J}_y.
\label{eq:TCM approx H}
} This approximation is valid for $\bar{n}-J+1/2\gg 1$ and all of the interaction times we consider here $(\Omega_0 t\sim \bar{n}^{-1/2}\ll \bar{n})$). It serves to rotate the collective atomic state at a Rabi frequency
\eq{
\Omega(J,n)=\Omega_0\sqrt{n-J+1/2}
} for a given field-state energy level $\ket{n}$, which is smaller than $\Omega_n$ above and decreases with $J$, explaining why slightly longer interaction times are required with increasing $J$ to achieve the same $\tfrac{\pi}{2}$ pulses [Fig. \ref{fig:TCM optimals}\textbf{(c)}]. However, the actual functional dependence in Fig. \ref{fig:TCM optimals}\textbf{(c)} looks like it follows $\Omega(J/2,n)$ instead of $\Omega(J,n)$, so we will investigate further to elucidate whether this is simply a numerical artifact. That one cannot simply replace $\bar{n}\to\bar{n}-J+1/2$ in Eqs. \eqref{eq:optimal params TCM} to match the replacement $\Omega_n\to\Omega(J,n)$ may be justified by the competition between Rabi frequencies with a variety of values of $n$.

We can approximate the full evolution of our initial state using Eq. \eqref{eq:TCM approx H} and the unitary evolution
\eq{
U_{\mathrm{TC}}=Q\exp\left(-\iu\tilde{H}_{\mathrm{TC}}t\right)Q^\dagger.
} Here, 
\eq{
Q=\sum_{m=-J}^J\eu^{\iu\hat{\phi}(J+m)}\otimes\ket{J,m}\bra{J,m}
} is an almost-unitary operator ($\left[Q,Q^\dagger\right]=|0\rangle\langle 0|$) and \eq{
\exp(\iu \hat{\phi})=\sum_{n=0}^\infty \ket{n}\bra{n+1}
} is a phase operator for the field (we have reserved the caret for the operator $\hat{\phi}$ in deference to the intricacies of phase operators \cite{BarnettVaccaro2007}). The first transformation leaves the state unchanged: 
\eq{
Q^\dagger \sum_n \psi_n\ket{n}\otimes \ket{J,-J}=\sum_n \psi_n\ket{n}\otimes\ket{J,-J}.
}
Next, the effective Hamiltonian enacts a rotation of the atomic states by an angle that depends on the field's energy level in a manner reminiscent of Eq. \eqref{eq:pi/2 pulse on 2J atoms}:
\eq{
\eu^{-\iu\tilde{H}_{\mathrm{TC}}t}\sum_n \psi_n\ket{n}\otimes\ket{J,-J}=&\sum_n \psi_n\ket{n}\\
&\otimes \sum_{m=-J}^J\sqrt{\binom{2J}{m+J}}\cos^{J-m}\frac{\Omega(J,n)t}{2}\sin^{J+m}\frac{\Omega(J,n)t}{2}\ket{J,m}.
\label{eq:midway approximate psi TCM}
}
Note that the atomic states in the superposition are SU(2)-coherent states that are eigenvalues of the spin operator pointing at different angles for different field energy levels $J_x \sin\left[\Omega(J,n)t\right]-J_z\cos\left[\Omega(J,n)t\right]$; if this was the end of the evolution, an initial field state with definite photon number $n$ would suffice to perfectly enact a rotation by $\Omega(J,n)$ on all of the atoms.
Finally, using
\eq{
\exp(\iu \hat{\phi}k)=\sum_{n=0}^\infty \ket{n}\bra{n+k},
} the transformed state becomes
\eq{
\ket{\Psi(t)}=&\sum_{n=2J+1}^\infty\sum_{m=-J}^J 
\sqrt{\binom{2J}{m+J}}\psi_{n+J+m}\ket{n}\otimes\ket{J,m}\\
&\times \cos^{J-m}\frac{\Omega(J,n+J+m)t}{2}\sin^{J+m}\frac{\Omega(J,n+J+m)t}{2},
\label{eq:approx psi(t) TCM}
} where we have restricted our attention to states with $\psi_n= 0$ for $n\leq 2J$ such that $Q$ is unitary.
We can then inspect the properties of this state to see how to optimally achieve the $\tfrac{\pi}{2}$ pulse of Eq. \eqref{eq:pi/2 pulse on 2J atoms}.

\begin{figure}
    \centering
    \includegraphics[width=0.65\columnwidth]{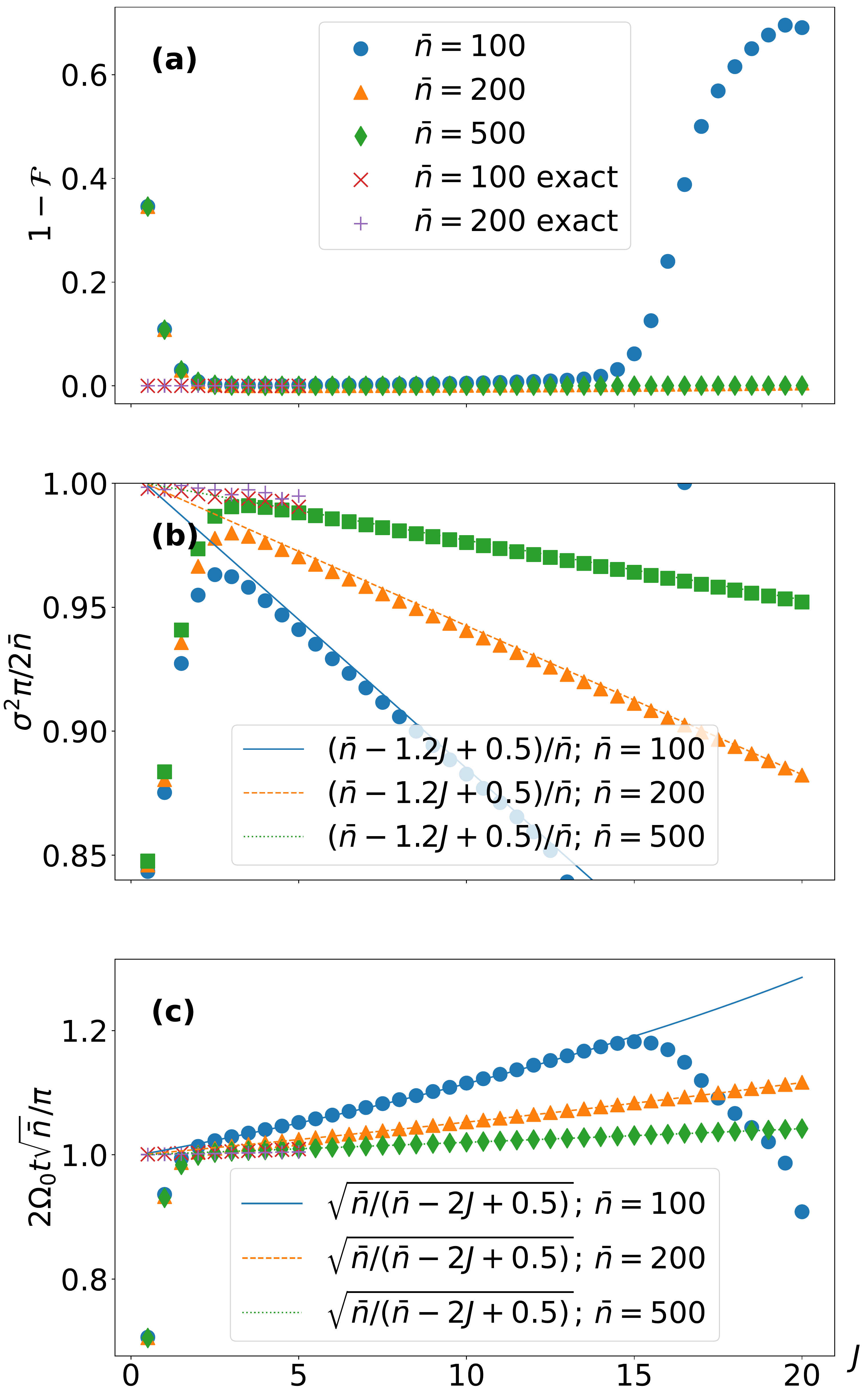}
    \caption{Optimal $\tfrac{\pi}{2}$ pulses on a collection of $2J$ atoms as in Fig. \ref{fig:TCM optimals}, but now optimized using the approximate fidelity of Eq. \eqref{eq:approx fidelity TCM} to permit larger $\bar{n}$ and $J$ ($\bar{n}=100$, $200$, and $500$ are the blue dots, orange triangles, and green diamonds, respectively). The approximation breaks down when $\sqrt{\bar{n}}\gg J$ is not achieved. \textbf{(a)} The error probabilities $1-\mathcal{F}$ with this method are again quite low but are nonnegligible when the approximation fails. \textbf{(b)} The optimal variances best match the curve $\sigma^2=2(\bar{n}-1.2J+0.5)/\pi$. \textbf{(c)} The optimal interaction times best match the curve $\Omega_0 t\sqrt{\bar{n}-2J+0.5}=$$\tfrac{\pi}{2}$. In all cases, the solutions found using the exact evolution (red $\times$s and purple $+$s for $\bar{n}=100$ and $200$, respectively) best match the $\bar{n}=500$ solution (i.e., the largest energy considered) for all $J$.
    }
    \label{fig:approx infidelities}
\end{figure}

For Eq. \eqref{eq:approx psi(t) TCM} to best approximate a $\tfrac{\pi}{2}$ pulse, we would like
\eq{
\psi_{n+k} \cos^{2J-k}\frac{\Omega(J,n+k)t}{2}\sin^{k}\frac{\Omega(J,n+k)t}{2}\approx \frac{1}{2^J}
} for all values of $n$ and $k$ such that the final state is most separable and the atomic state is closest to that of Eq. \eqref{eq:pi/2 pulse on 2J atoms}. The width of the trigonometric terms' distribution changes with $n$ and $k$, so it is not obvious how to choose the appropriate optimal width for the photon-number distribution, although we note that all Gaussian distributions centred at $\bar{n}$ with interaction times $\Omega(J,\bar{n})t\approx$$\tfrac{\pi}{2}$ will converge to the proper limit with large $\bar{n}$. Instead, we can look at the overlap between $\ket{\Psi(t)}$ and a state rotated by $\Theta$:
\eq{
\left|\braket{J,\Theta}{\Psi(t)}\right|^2=
\sum_{n=2J+1}^\infty\left|\sum_{m=-J}^{J} 
{\binom{2J}{m+J}}\psi_{n+J+m}\cos^{J-m}\frac{\Omega(J,n+J+m)t}{2}\right.\\
\sin^{J+m}\frac{\Omega(J,n+J+m)t}{2}
\left.\times \cos^{J-m}\frac{\Theta}{2}\sin^{J+m}\frac{\Theta}{2}\right|^2.
\label{eq:approx fidelity TCM}
} By selecting $\Theta=$$\tfrac{\pi}{2}$ and Gaussian photon-number distributions centred at $\bar{n}$, we can optimize this result over all interaction times $\Omega_0 t$ and photon-number variances $\sigma^2$. Exemplary results are plotted in 
Fig. \ref{fig:approx infidelities}, with the data from Fig. \ref{fig:TCM optimals} overlain. 

Comparing the results between the two optimization methods in Fig. \ref{fig:approx infidelities}, we see that they match in the case of large $\bar{n}$. Intriguingly, the replacement $\bar{n}\to \bar{n}-qJ+1/2$ in Eq. \eqref{eq:optimal params TCM} seems to always hold, but the optimal order-unity parameter $q$ does not seem consistent between $\sigma^2_{\mathrm{optimal}}$ and $t_{\mathrm{optimal}}$. 
This may be partially explained by setting $|\psi_{\bar{n}+\delta}|\propto \exp(-\delta^2/4\sigma^2)$ for small $\delta$ in Eq. \eqref{eq:approx fidelity TCM}, expanding around $n=\bar{n}$, expanding the sinusoidal terms around large $\bar{n}$, and asking what photon-number variance $\sigma^2$ will cancel all of the $\mathcal{O}(\delta)$ terms; the result is
\eq{
\sigma^2
=\frac{2J+m+m^\prime}{m+m^\prime}\left[\frac{2(\bar{n}-J+1/2)}{\pi}+\frac{J (m+m^\prime)+m^2+m^{\prime 2}}{2}\right],
} which resembles Eq. \eqref{eq:optimal params TCM} with the replacement $\bar{n}\to\bar{n}-J+1/2$ but changes with different values of $m$ and makes no sense (goes negative) when $m+m^\prime\leq 0$.
It is thus always a good approximation to use Eq. \eqref{eq:optimal params TCM} and then update $\bar{n}$ as a function of $J$ for the problem at hand.

A complementary strategy for optimizing the initial field states is to look at the evolution of the expectation values of $J_z$ and $J_x$. For a perfect $\tfrac{\pi}{2}$ pulse, the expectation value of $J_z$ should go from its minimal value $-J$ to $0$, while that of $J_x$ should go from $0$ to its maximal value $J$. In fact, given a fixed total spin, any of the collective operators $J_i$ attaining its maximal eigenvalue is a sufficient condition for the spin state to be in a pure state and thus for there to be no residual entanglement with the light. These goals can then give us constraints on our initial field states in the spirit of transcoherence.

The $J_z$ operator evolves in the Heisenberg picture as \cite{KlimovChumakov1995}
\eq{
U_{\mathrm{TC}}J_z U_{\mathrm{TC}}^\dagger=J_z \cos\hat{\tau}-J_y\sin\hat{\tau}
} for the field operator $\hat{\tau}=\Omega_0 t\sqrt{\had\ha-J+1/2}$. Since the initial atomic state has \aaron{expectation values} $\expct{(J_x,J_y,J_z)}=(0,0,-J)$, the final state has
\eq{
\langle \Psi(t)|J_z|\Psi(t)\rangle=-J\expct{\cos\hat{\tau}},
} where the final expectation value is taken with respect to the initial field state. For this quantity
\eq{
\expct{\cos\hat{\tau}}=\sum_n |\psi_n|^2\cos\left(\Omega_0 t\sqrt{n-J+1/2}\right)
}
to vanish, the photon-number distribution should be strongly peaked around $\expct{\tau}=$$\tfrac{\pi}{2}$, again corresponding to a classical $\tfrac{\pi}{2}$ pulse with average field strength $\bar{n}$ and interaction time $\Omega(J,\bar{n})t=$$\tfrac{\pi}{2}$.

We then look at the evolution of $J_x$. Since we already have an expression for the evolved state, we can directly calculate
\eq{
\langle \Psi(t)|J_x|\Psi(t)\rangle=\sum_{n=2J+1}^\infty\sum_{m=-J}^J\binom{2J}{J+m}(J-m)\psi_{n+J+m}\psi_{n+J+m+1}^* \cos^{J-m}\frac{\Omega(J,n+J+m)t}{2}\\
\times\sin^{J+m}\frac{\Omega(J,n+J+m)t}{2}%\\
%\times 
\cos^{J-m-1}\frac{\Omega(J,n+J+m+1)t}{2}\sin^{J+m+1}\frac{\Omega(J,n+J+m+1)t}{2}.
} Expanding around $n+J=\bar{n}+\delta$ for small $\delta$ and again using $|\psi_{\bar{n}+\delta}|\propto\exp(-\delta^2/4\sigma^2)$, we find
\eq{
\frac{\psi_{n+J+m}\psi_{n+J+m+1}^*}{|\psi_{\bar{n}}|^2}\approx 1-\frac{(\delta+m)^2+(\delta+m+1)^2}{4 \sigma^2}+\mathcal{O}(\sigma^{-4}).
} The sinusoidal terms expand to leading order in $\bar{n}-J+1/2$ as 
\eq{
4^{-J}\left\{1+\frac{\pi  \left[2 \delta m+\delta+2 m (m+1)+1\right]}{4\left(\bar{n}-J+1/2\right)}\right\}.%+\mathcal{O}\left[\frac{1}{\left(\bar{n}-J+1/2\right)^2}\right]
\nonumber
} Multiplying these terms by the coefficients $\binom{2J}{J+m}(J-m)$ and summing from $m=-J$ to $m=J$ yields, to leading order in $\sigma^2$ and $\bar{n}-J+1/2$,
\eq{
\langle \Psi(t)|J_x|\Psi(t)\rangle\approx J\sum_{\delta} |\psi_{\bar{n}}|^2\left[ 1-\frac{ 2 \delta^2+J}{4 \sigma^2}+\frac{\pi J}{4 (\bar{n}-J+1/2)}\right].
} Performing the sum from $\delta=-l$ to $\delta=l$ for some $l$, the leading-order terms cancel each other when
\eq{
\sigma^2=\frac{\bar{n}-J+1/2}{\pi }\left(1+\frac{2l(l+1)}{3J}\right).
} This is exactly the result of Eq. \eqref{eq:optimal params TCM} with the replacement $\bar{n}\to\bar{n}-J+1/2$ and the number-squeezing factor being replaced as $\tfrac{2}{\pi}\to \left(1+\tfrac{2l(l+1)}{3J}\right)/\pi$; if the normalization is given by $|\psi_{\bar{n}}|^2\approx 1/\sqrt{2J+1}$, we find $l=(\sqrt{2J+1}-1)/2$ and the number-squeezing factor being exactly $\tfrac{2}{\pi}$ together yield the best result for $\langle \Psi(t)|J_x|\Psi(t)\rangle\approx J$. The overall dependence of the optimal variance $\sigma^2$ on $\bar{n}$ and $J$ is now apparent and considerations of different ranges of the sum over $\delta$ change the dependence of the optimal variance on $J$.

\subsection{Perfect pulses cannot be generated for the Tavis-Cummings interaction}~
The evolution of the TCM can always be solved exactly \cite{TavisCummings1968,Scharf1970,HeppLieb1973} but requires the intricate solution of the Bethe ansatz equations \cite{Bogoliubovetal1996,Luetal2021}. We solve this system of equations for the case of $N=2$ atoms, given the initial state $\sum_n \psi_n\ket{n}\otimes \ket{J,-J}$, to show that perfect $\tfrac{\pi}{2}$ pulses are not generally possible. Since total excitation number is conserved and the atomic subspace is spanned by only three states $\ket{1,\pm 1}$ and $\ket{1,0}$, we can analytically solve this model. 

The evolved state at any time $t$ is (using $\ket{m,n}\equiv\ket{m}\otimes\ket{J,n}$ for brevity in this subsection alone):
\eq{
&\ket{\Psi(t)}=\psi_0\ket{0,-1}+\psi_1\left(\cos\frac{\Omega_0 t}{\sqrt{2}}\ket{1,-1}-\iu\sin\frac{\Omega_0 t}{\sqrt{2}}\ket{0,0}\right)\\
&+\sum_{n\geq 2} \psi_n\ket{n,-1}\left(\frac{n-1}{2n-1}+\frac{n}{2n-1}\cos\frac{\Omega_0 t\sqrt{2n-1}}{\sqrt{2}}\right)\\
&+ \psi_n\ket{n-2,1}\frac{\sqrt{n(n-1)}}{2n-1}\left(-1+\cos\frac{\Omega_0 t\sqrt{2n-1}}{\sqrt{2}}\right)-\iu \psi_n\ket{n-1,0}\sqrt{\frac{n}{2n-1}}\sin\frac{\Omega_0 t\sqrt{2n-1}}{\sqrt{2}}.
}
We found this using the eigenstates 
\eq{
\ket{n\pm}=\frac{1}{\sqrt{2}}\left(\sqrt{\frac{n}{2n-1}}\ket{n,-1}\pm\ket{n-1,0}+\sqrt{\frac{n-1}{2n-1}}\ket{n-2,1}\right)
} with eigenvalues $\pm\sqrt{2n-1}\Omega_0/\sqrt{2}$ and the null eigenstate
\eq{
\ket{n0}=-\sqrt{\frac{n-1}{2n-1}}\ket{n,-1}+\sqrt{\frac{n-1}{2n-1}}\ket{n-2,1}.
} Can this ever perfectly create the desired pulse?

Projecting onto $\ket{0}$ for the field state, the first requirement for the atoms to be in the correct rotated state is that
\eq{
\psi_0\ket{-1}-\iu \psi_1\sin\frac{\Omega_0 t}{\sqrt{2}}\ket{0}+\psi_2\frac{\sqrt{2}}{3}\left(\cos\frac{\Omega_0 t\sqrt{3}}{\sqrt{2}}-1\right)\ket{1}
\propto \ket{-1}+\sqrt{2}\ket{0}+\ket{1}.
} This immediately yields two constraints for the three free parameters $\psi_1$, $\psi_2$, and $t$:
\eq{
\psi_0&=\psi_2\frac{\sqrt{2}}{3}\left(\cos\frac{\Omega_0 t \sqrt{3}}{\sqrt{2}}-1\right)\aaron{,}\\
\psi_0\sqrt{2}&=-\iu \psi_1\sin\frac{\Omega_0 t}{\sqrt{2}}.
} If any of these three coefficients vanishes then they all must, unless the timing spares $\psi_1$ or $\psi_2$ from needing to vanish. The next requirement found by projecting the field onto state $\ket{1}$ is that
\eq{
&\psi_1\cos\frac{\Omega_0 t}{\sqrt{2}}\ket{-1}-\iu \psi_2\sqrt{\frac{2}{3}}\sin\frac{\Omega_0 t\sqrt{3}}{\sqrt{2}}\ket{0}
+\psi_3\frac{\sqrt{6}}{5}\left(\cos\frac{\Omega_0 t\sqrt{5}}{\sqrt{2}}-1\right)\ket{1}
\propto \ket{-1}+\sqrt{2}\ket{0}+\ket{1}.
} Again, if any of the coefficients vanishes then they all must, unless rescued by an exact timing prescription. This introduces another required relationship between $\psi_1$ and $\psi_2$,
\eq{
\psi_1\sqrt{2}\cos\frac{\Omega_0 t}{\sqrt{2}}=-\iu \psi_2\sqrt{\frac{2}{3}}\sin\frac{\Omega_0 t\sqrt{3}}{\sqrt{2}},
}
which together impose the timing requirement:
\eq{
\frac{\frac{2}{3}\left(\cos\frac{\Omega_0 t \sqrt{3}}{\sqrt{2}}-1\right)}{-\iu\sin\frac{\Omega_0 t}{\sqrt{2}}}&=
-\iu \frac{\frac{1}{\sqrt{3}}\sin\frac{\Omega_0 t \sqrt{3}}{\sqrt{2}}}{\cos\frac{\Omega_0 t}{\sqrt{2}}}\\
\aaron{\Rightarrow}\quad \frac{2}{\sqrt{3}}\left(\cos\frac{\Omega_0 t \sqrt{3}}{\sqrt{2}}-1\right)&=-\sin\frac{\Omega_0 t \sqrt{3}}{\sqrt{2}}
\tan\frac{\Omega_0 t}{\sqrt{2}}.
} This only holds when $\Omega_0 t \sqrt{3/2}=2k\pi,\,k\in\mathds{N}$. But then we find that $\psi_0=0$, $\psi_1=0$, and so on, and we have no solution. So it is impossible to perfectly solve this problem for two atoms. %We should not expect it to be exactly solvable in general. Ok, but can it be solved for these coefficients being zero and using another set of coefficients that don't start from $\psi_0$? Now there are some additional time points that are good for adjacent number states that don't require things vanishing. But they cannot be satisfied for more than one set of adjacent things at once, so it will never work, because even a pair of adjacent things needs each of the two edges to also work, and they can't vanish, but then the next space at their edges won't be satisfied!

Even considering the general case where the transformed state approximately follows Eq. \eqref{eq:approx psi(t) TCM}, the parameters cannot be chosen precisely enough such that no excitation escapes. Consider that there will always be some probability of finding the atom in any state $\ket{J,m}$ for rotations with $0<\Theta<\pi$. Every field state $\ket{n}$ must be in a tensor product with the same superposition of atomic states, so, whenever any coefficient $\psi_{n+J+m}$ is nonzero, every other coefficient $\psi_{n+J+m^\prime}$ must also be nonzero for all $-J\leq m^\prime\leq J$. Since $n$ can vary by $1$ and the range of values of $m$ must vary by at least $2$ for any $J>1/2$ (i.e., for anything but the JCM case of a single atom), there can be no maximal coefficient beyond which all of the coefficients vanish. Some excitation will always leak out of the subspaces in which the atoms are in the proper rotated state. The sole alternative to setting coefficients to zero is that one of the sinusoidal terms vanishes for a particular $n$ and $m$ and $t$. This suffices in the case of the JCM because there is only a single maximal coefficient that must be constrained, but is insufficient in the $J>1/2$ case where entire ranges of coefficients must vanish in the field state's maximal photon number. This is why, although squeezing is beneficial regardless of $J$, the true transcoherent states that perfectly enact $\Theta$ rotations only exist for the case of $J=1/2$. We note incidentally that a true transcoherent state for arbitrary $J$ can be achieved in the trivial case of a rotation by $0$ (i.e., the field should be in the vacuum state) and for $\Theta=\pi$ with a Fock state. This is exact for the JCM and holds precisely within the approximations of the TCM that lead to Eq. \eqref{eq:approx psi(t) TCM}, because the state in Eq. \eqref{eq:midway approximate psi TCM} is separable and remains separable following the application of the almost-unitary operator $Q^\dagger$ on the state.

\subsection{Discussion}~
Photon-number squeezing increases the probability of successfully imparting a $\tfrac{\pi}{2}$ pulse on a collection of ground-state atoms. Given the ubiquity of this result for other initial atomic states in the JCM, we expect photon-number squeezing to similarly enhance arbitrary pulse areas in the TCM for unknown initial atomic states. This problem is slightly more numerically cumbersome due to the lack of a closed-form expression for the fidelities averaged over all initial atomic states, so we leave this expectation as a conjecture that could be evidenced by numerical investigations of Eq. \eqref{eq:approx fidelity TCM} with various $\Theta$.

Earlier studies showed that a collection of partially excited atoms in the final state of Eq. \eqref{eq:pi/2 pulse on 2J atoms} will interact with a coherent field state to number squeeze the latter \cite{Retamaletal1997}. It thus comes as no surprise that photon-number-squeezed states are useful for enacting, in some sense, the reverse of this process. This idea of matching the squeezing to the interplay between different Rabi frequencies should be useful for a variety of light-matter-interaction protocols.

\section{Arbitrary coherence cannot be generated in the presence of nonzero detuning}~
All of the models thus far have dealt with resonant interactions between the field mode and the atoms. We show here that the same can \textit{never} be achieved \textit{exactly} for nonzero values of detuning. 

One reason to consider nonzero detuning is to establish transformations like the Hadamard transform. This sends an atom in its ground state to an even superposition of its ground and excited state, but does so by a $\pi$ rotation of the spin vector about the $\tfrac{x+z}{\sqrt{2}}$-axis of the Bloch sphere, instead of the $\tfrac{\pi}{2}$ pulses we have been discussing here. For large field states, the interaction parts of the JCM and TCM Hamiltonian effectively rotate the average spin vector about some axis in the $xy$-plane, so the only method for truly imparting a Hadamard gate is by the introduction of a nonzero detuning that allows for effective rotations about other spin axes.

The Jaynes-Cummings interaction Hamiltonian at nonzero detuning $\delta$ takes the form
\eq{
H_{\mathrm{II}}=\frac{\delta}{2}\sigma_z+\frac{\Omega_0}{2}\left(\ha\sigma_-+\had\sigma_+\right).
} The new eigenstates are
\eq{
\ket{+,n}_\delta&=\cos\frac{\alpha_n}{2}\ket{n}\otimes\ket{\mathrm{e}}+\sin\frac{\alpha_n}{2}\ket{n+1}\otimes\ket{\mathrm{g}},\\
\ket{-,n}_\delta&=\sin\frac{\alpha_n}{2}\ket{n}\otimes\ket{\mathrm{e}}-\cos\frac{\alpha_n}{2}\ket{n+1}\otimes\ket{\mathrm{g}}
} and have interaction-picture energies
\eq{
E_\pm(n)=\pm\frac{\Omega(n)}{2},
}
where \eq{
\alpha_n=\tan^{-1}\frac{\Omega_0\sqrt{n+1}}{\delta}
} and we have defined the detuned Rabi frequencies
\eq{
\Omega(n)=\sqrt{\Omega_n^2+\delta^2}=\sqrt{\Omega_0^2(n+1)+\delta^2}.
} An atom initially in its ground state and the field initially in a general state $\sum_n \psi_n\ket{n}$ together evolve to
\eq{
\ket{\Psi(t)}=&\sum_{n=-1}^\infty \psi_{n+1}\left[
-\iu \sin\frac{\Omega(n)t}{2}\sin\frac{\alpha_n}{2}\cos\frac{\alpha_n}{2}\ket{n}\otimes\ket{\mathrm{e}}\right.\\
&\left.
+\left(\eu^{-\iu\Omega(n)t/2}\sin^2\frac{\alpha_n}{2}+\eu^{\iu\Omega(n)t/2}\cos^2\frac{\alpha_n}{2}\right)\ket{n+1}\otimes\ket{\mathrm{g}}\right].
}

We can proceed as usual to find initial field states and interaction times that create separable states with the atom completely coherent. Here, there are more complicated conditions that need to be solved, but there is the extra degree of freedom in the detuning that could account for them.

Projecting the evolved state onto states with definite photon number and requiring the result to be proportional to $\cos\frac{\theta}{2}\ket{\mathrm{g}}+\sin\frac{\theta}{2}\ket{\mathrm{e}}$ leads to the recursion relation:
\eq{
\frac{\psi_{n+1}}{\psi_n}=-\iu\tan\frac{\theta}{2}%\left(\eu^{-\iu\Omega(n-1)t/2}\sin^2\frac{\alpha_{n-1}}{2}+\eu^{\iu\Omega(n-1)t/2}\cos^2\frac{\alpha_{n-1}}{2}\right)
\frac{\eu^{-\iu\Omega(n-1)t/2}\sin^2\frac{\alpha_{n-1}}{2}+\eu^{\iu\Omega(n-1)t/2}\cos^2\frac{\alpha_{n-1}}{2}}{\sin\frac{\Omega(n) t}{2}\sin\frac{\alpha_n}{2}\cos\frac{\alpha_n}{2}}.
} For a given $t$, $\Omega_0$, and $\delta$, this series is uniquely determined. Can we force this series to truncate on both sides?

To not have any probability leak out of the lowest-excitation subspace with the smallest nonzero coefficient $\psi_{n_{\mathrm{min}}}$, we require
\eq{
\sin\frac{\Omega({n_{\mathrm{min}}-1}) t}{2}\sin\frac{\alpha_{n_{\mathrm{min}}-1}}{2}\cos\frac{\alpha_{n_{\mathrm{min}}-1}}{2}=0.
}
This constraint is readily satisfied when $n_{\mathrm{min}}=0$, because $\alpha_{-1}=0$.
For larger $n_{\mathrm{min}}$, this amounts to the requirement
\eq{
\Omega({n_{\mathrm{min}}-1}) t=\Omega_0 t\sqrt{ n_{\mathrm{min}}+\left(\frac{\delta}{\Omega_0}\right)^2}=2m\pi,\quad m\in\mathds{N},
}
which can readily be satisfied by an appropriate interaction time $\Omega_0 t$ and detuning $\delta$. 

However, it is impossible to not have any probability leak out of the highest-excitation subspace with the largest nonzero coefficient $\psi_{n_{\mathrm{max}}}$. To do so, we would require both
\eq{
\cos\frac{\Omega({n_{\mathrm{max}}}-1)t}{2}\left(\sin^2\frac{\alpha_{{n_{\mathrm{max}}}-1}}{2}+\cos^2\frac{\alpha_{{n_{\mathrm{max}}}-1}}{2}\right)=0
} and
\eq{
\sin\frac{\Omega({n_{\mathrm{max}}}-1)t}{2}\left(\sin^2\frac{\alpha_{{n_{\mathrm{max}}}-1}}{2}-\cos^2\frac{\alpha_{{n_{\mathrm{max}}}-1}}{2}\right)=0.
} While the former is readily converted into the satisfiable condition
\eq{
\sqrt{\Omega_0^2 n_{\mathrm{max}}+\delta^2}t&=(2l+1)\pi,\quad l\in\mathds{N}
,
}
the latter can \textit{only} be satisfied by \eq{
\alpha_{n_{\mathrm{max}}-1}=\frac{\pi}{2} \quad\Rightarrow\quad \delta=0.
} Any presence of nonzero detuning allows excitations to leak beyond the highest-excitation subspace, so there can never be perfect transfer of coherence from a field state to an atom that leaves no residual entanglement when the interaction is nonresonant. This accords with complete transfer of probability between $\ket{\mathrm{g}}$ and $\ket{\mathrm{e}}$ being impossible with nonzero detuning in the Rabi model with a classical field.

\section{$m$-photon processes}~
\subsection{Beyond the Jaynes-Cummings model}~
What happens when it takes more than one photon to excite an atom? We can consider a nonlinear interaction that requires $m$-photon absorption to transform $\ket{\mathrm{g}}$ to $\ket{\mathrm{e}}$:
\eq{
H=\omega\left(\frac{1}{m}\had\ha+\ket{\mathrm{e}}\bra{\mathrm{e}}\right)+\frac{\Omega_0^{(m)}}{2}\left(\ha^m\sigma_+ + \had^m\sigma_-\right),
} where $\omega$ is the resonance frequency but now each individual photon provides energy $\tfrac{\omega}{m}$ and $\Omega_0^{(m)}$ is the coupling strength that depends on the $m$th-order nonlinearity. This interaction conserves total energy and a form of the total excitation number, as can be seen from its eigenstates
\eq{
\ket{\pm,n}=\frac{\ket{n}\otimes\ket{\mathrm{e}}\pm\ket{n+m}\otimes\ket{\mathrm{g}}}{\sqrt{2}},
%\label{eq:JCM eigenstates}
} which now have the quantized-Rabi-like frequencies
\eq{
\Omega_n^{(m)}=\Omega_0^{(m)}\sqrt{(n+m)(n+m-1)\cdots(n+1)}.
} We will work in the interaction picture with Hamiltonian
\eq{
H_{\mathrm{I}}=
\frac{\Omega_0^{(m)}}{2}\left(\ha^m\sigma_+ + \had^m\sigma_-\right);
} the Schr\"odinger-picture results can thence be obtained with the substitutions $\ket{n}\to\eu^{-\iu\omega nt/m}\ket{n}$ and $\ket{\mathrm{e}}\to\eu^{-\iu \omega t}\ket{\mathrm{e}}$.

When the atom is initially in its ground state and the field in state $\sum_n \psi_n\ket{n}$, the evolved state takes the form [c.f. Eq. \eqref{eq:JCM from ground}]
\eq{
\ket{\Psi(t)}
=&\sum_{n=0}^\infty \ket{n}\otimes\left(\psi_n\cos\frac{\Omega_{n-m}^{(m)}t}{2}\ket{\mathrm{g}}-\iu \psi_{n+m}\sin\frac{\Omega_n^{(m)} t}{2}\ket{\mathrm{e}}\right).
} Similarly,
when the atom is initially in its excited state and the field in state $\sum_n \psi_n\ket{n}$, the evolved state takes the form [c.f. Eq. \eqref{eq:JCM from excited}]
\eq{
\ket{\Psi(t)}=&\sum_{n=0}^\infty \ket{n}\otimes \left(\psi_n\cos\frac{\Omega_n^{(m)} t}{2}\ket{\mathrm{e}}
-\iu \psi_{n-m}\sin\frac{\Omega_{n-m}^{(m)} t}{2}\ket{\mathrm{g}}\right).
}
\aaron{We have again abused notation in the definition of the new type of Rabi frequencies to set $\Omega_{k}^{(m)}$=0 for $k<0$.}

\subsection{Transcoherent states and beyond}~
What are the optimal field states that can generate arbitrary pulse areas for this nonlinear interaction? 
We again seek transformations for which the final state has zero residual entanglement between the atom and the light, such that the atomic state can be used in arbitrary quantum information protocols without degradation.

From the ground state, arbitrary transformations can be performed by field states whose photon-number coefficients satisfy
\eq{
\psi_{n+m}=\iu\tan\frac{\theta}{2}\frac{\cos\frac{\Omega_{n-m}^{(m)} t}{2}}{\sin\frac{\Omega_n^{(m)} t}{2}}\psi_n
} to ensure that the amplitudes of $\ket{\mathrm{g}}$ and $\ket{\mathrm{e}}$ in the evolved state in Eq. \eqref{eq:JCM from ground} are equal. This can be satisfied by field states with $\psi_n=0$ for $n> n_{\mathrm{max}}$ for some chosen $n_{\mathrm{max}}\geq 1$, so long as the total interaction time satisfies
\eq{
\Omega_{n_{\mathrm{max}}-1}^{(m)}t=\pi,
} which ensures that the highest-excitation subspace spanned by $\ket{\pm,n_{\mathrm{max}}-1}$ undergoes a $\pi$ pulse. Now, in contrast to the $m=1$ scenario, there are $m$ independent recursion relations that must all truncate at the same time $t$, which cannot occur because the oscillation frequencies cannot have an integer ratio $\Omega_{n_k}^{(m)}/\Omega_{n}^{(m)}$ for any integer $k$. Therefore, in order to exactly produce the desired atomic state, one must use a state of light that sets to zero $m-1$ of the coefficients from $\psi_0$ to $\psi_{m-1}$ and thus that only has population in photon numbers spaced $m$ apart. The alternative, which can still outperform coherent states, is to use a state with a large average number of photons that will approximate the recursion relation by a squeezed state.

The same can be said for the atom initially in its excited state, where now the optimal recursion relation takes the form
\eq{
 \psi_{n+m}=-\iu\tan\frac{\theta}{2}\frac{\sin\frac{\Omega_{n}^{(m)} t}{2}}{\cos\frac{\Omega_{n+m}^{(m)} t}{2}}\psi_{n}.
} 

What are the properties of the field states that approximate these recursion relations in the limit of large numbers of photons? As usual, the optimal interaction time matches the semiclassical one:
\eq{
\Omega_0^{(m)} t&=\frac{\theta}{\sqrt{(\bar{n}+m)(\bar{n}+m-1)\cdots(\bar{n}+1)}}
\approx \frac{\theta}{\sqrt{\bar{n}^m+\frac{m(m+1)}{2}\bar{n}^{m-1}}}
} from the ground state and 
\eq{
\Omega_0^{(m)} t\approx\frac{\theta+\pi}{\sqrt{\bar{n}^m+\frac{m(m+1)}{2}\bar{n}^{m-1}}}
} from the excited state. The ratios in the recursion relations obey
\eq{
\tan\frac{\theta}{2}\frac{\cos\frac{\Omega_{\bar{n}+\delta-m}^{(m)} t}{2}}{\sin\frac{\Omega_{\bar{n}+\delta}^{(m)} t}{2}}\approx 1-\frac{m  }{2 \bar{n}\sinc\theta}\left(\delta-m\sin^2\frac{\theta}{2}\right)
} and (recall that starting in $\ket{\mathrm{e}}$ requires $\sinc(\theta+\pi)$ to be negative in order to rotate by $\theta+\pi$ to $\ket{\theta}$)
\eq{
 -\tan\frac{\theta}{2}\frac{\sin\frac{\Omega_{\bar{n}+\delta}^{(m)} t}{2}}{\cos\frac{\Omega_{\bar{n}+\delta+m}^{(m)} t}{2}}\approx 1+\frac{m}{2\bar{n}\sinc(\theta+\pi)}\left(\delta+m\sin^2\frac{\theta+\pi}{2}\right).
} For comparison, an exponential distribution for coefficients separated by $m$ obeys
\eq{
\left|\frac{\psi_{\bar{n}+\delta+m}}{\psi_{\bar{n}+\delta}}\right|=\exp\left[-\frac{(\delta+m)^2-\delta^2}{4\sigma^2}\right]\approx1-\frac{m}{2\sigma^2}\left(\delta+\frac{m}{2}\right).
} We see that \textit{no different number squeezing is needed} for $m$-photon processes relative to the JCM, with the mean photon numbers being shifted from the ideal classical ones by a factor of $\pm m\sin^2\frac{\theta}{2}$.
\color{black}

\section{Conclusions}~
We have performed a detailed investigation of the optimal field states for transferring arbitrary amounts of coherence to individual and collections of atoms. When a single atom is initially its ground or excited state, there exists a field state to rotate it by an arbitrary amount in an arbitrarily short amount of time that generates no residual entanglement with the field. Since these unitary operations can be reversed and, therefore, composed, we have thus found field states for perfectly performing arbitrary rotations on arbitrary atomic states without resorting to any semiclassical approximations.

The perfect field states and interaction times depend on knowing the initial state of the atom. When this initial state is unknown, field states with their photon-number distributions squeezed relative to coherent states retain an advantage in their ability to perform arbitrary operations on some average atomic state. These squeezed field states can then be useful for tasks like creating logic gates for quantum computers.

We showed that squeezed light is also useful for transferring coherence to a collection of atoms or to any spin system. This cements squeezed light as a resource beyond traditional realms such as metrology \cite{LIGO2011} and computation \cite{Madsenetal2022}. More squeezing is required to perform larger rotations on atomic states, so the continuing improvements in squeezing capabilities motivated by said traditional applications provides increasing benefit to our light-matter-interaction scenarios.

Finally, we found that generalizations of the JCM to nonlinear processes responsible for high-harmonic generation and $m$-photon absorption can also have transcoherent states and beyond. The optimal field states for rotation atoms by $\theta$ through nonlinear interactions are \textit{also} squeezed in their photon-number variances by a factor of $\sinc\theta$. All of the results from linear interactions thus extend \textit{mutatis mutandis} to nonlinear ones.

Transcoherent states and beyond stimulate many questions that may be explored in future work. Are these states easiest to generate in a cavity; if so, does the number of atoms in the cavity affect how the field states may enter the cavity? \aaron{How does dissipation from the cavity affect the coherence transfer; do states with smaller $\bar{n}$ perform better, or do faster interactions using states with large $\bar{n}$ prevail?} Can they be generated in an optomechanical system using phonons instead of photons as the bosonic mode? If one instead uses a beam of light travelling through free space, for which the JCM is no longer the exact model \cite{KiilerichMolmer2019,KiilerichMolmer2020}, how does squeezing affect coherence transfer to atoms? Does transcoherence lead to better design of field states in the presence of nonzero detuning between the atomic energy gap and the field frequency; can there still be an advantage in coherence transfer due to squeezing? Are there other interactions beyond the JCM and the generalizations considered here for which squeezing can confer additional advantages relative to coherent light? These exciting questions are but a fraction of what can now be studied.

Our previous work explored quantum catalysis as a particularly useful application of transcoherent states that generate perfect $\tfrac{\pi}{2}$ pulses to individual atoms. Now that the toolbox has been expanded to arbitrary rotations and arbitrary numbers of atoms, we strongly believe that our transcoherent states and beyond will be important to application in which light is used to precisely control atoms in any desired fashion.

\begin{acknowledgments}
    AZG and KH acknowledge that the NRC headquarters is located on the traditional unceded territory of the Algonquin Anishinaabe and Mohawk people. This work was supported by NSERC. AMS acknowledges support as a CIFAR Fellow. AZG thanks Andrei Klimov for useful discussions.
\end{acknowledgments}

%\bibliographystyle{quantum}
%\bibliography{Bibliography}

\begin{thebibliography}{10}

\bibitem{Shor1999}
Peter~W. Shor.
\newblock ``Polynomial-time algorithms for prime factorization and discrete
  logarithms on a quantum computer''.
\newblock \href{https://dx.doi.org/10.1137/S0036144598347011}{SIAM Review {\bf
  41}, 303--332}~(1999).

\bibitem{Dowling1998}
Jonathan~P. Dowling.
\newblock ``Correlated input-port, matter-wave interferometer: Quantum-noise
  limits to the atom-laser gyroscope''.
\newblock \href{https://dx.doi.org/10.1103/PhysRevA.57.4736}{Physical Review A
  {\bf 57}, 4736--4746}~(1998).

\bibitem{Bouwmeesteretal1997}
Dik Bouwmeester, Jian-Wei Pan, Klaus Mattle, Manfred Eibl, Harald Weinfurter,
  and Anton Zeilinger.
\newblock ``Experimental quantum teleportation''.
\newblock \href{https://dx.doi.org/10.1038/37539}{Nature {\bf 390},
  575}~(1997).

\bibitem{Aberg2006arxiv}
Johan {Aberg}.
\newblock ``{Quantifying Superposition}''~(2006).
\newblock
  \href{http://arxiv.org/abs/quant-ph/0612146}{arXiv:quant-ph/0612146}.

\bibitem{Baumgratzetal2014}
T.~Baumgratz, M.~Cramer, and M.~B. Plenio.
\newblock ``Quantifying coherence''.
\newblock \href{https://dx.doi.org/10.1103/PhysRevLett.113.140401}{Physical
  Review Letters {\bf 113}, 140401}~(2014).

\bibitem{LeviMintert2014}
Federico Levi and Florian Mintert.
\newblock ``A quantitative theory of coherent delocalization''.
\newblock \href{https://dx.doi.org/10.1088/1367-2630/16/3/033007}{New Journal
  of Physics {\bf 16}, 033007}~(2014).

\bibitem{WinterYang2016}
Andreas Winter and Dong Yang.
\newblock ``Operational resource theory of coherence''.
\newblock \href{https://dx.doi.org/10.1103/PhysRevLett.116.120404}{Physical
  Review Letters {\bf 116}, 120404}~(2016).

\bibitem{GoldbergSteinberg2020}
Aaron~Z. Goldberg and Aephraim~M. Steinberg.
\newblock ``Transcoherent states: Optical states for maximal generation of
  atomic coherence''.
\newblock \href{https://dx.doi.org/10.1103/PRXQuantum.1.020306}{Physical Review
  X Quantum {\bf 1}, 020306}~(2020).

\bibitem{Korzekwaetal2016}
Kamil Korzekwa, Matteo Lostaglio, Jonathan Oppenheim, and David Jennings.
\newblock ``The extraction of work from quantum coherence''.
\newblock \href{https://dx.doi.org/10.1088/1367-2630/18/2/023045}{New Journal
  of Physics {\bf 18}, 023045}~(2016).

\bibitem{Mulleretal2009}
M.~M\"uller, I.~Lesanovsky, H.~Weimer, H.~P. B\"uchler, and P.~Zoller.
\newblock ``Mesoscopic {R}ydberg gate based on electromagnetically induced
  transparency''.
\newblock \href{https://dx.doi.org/10.1103/PhysRevLett.102.170502}{Physical
  Review Letters {\bf 102}, 170502}~(2009).

\bibitem{AllenEberly1987}
Leslie Allen and Joseph~H Eberly.
\newblock ``Optical resonance and two-level atoms''.
\newblock Volume~28.
\newblock Courier Corporation. ~(1987).

\bibitem{Sudarshan1963}
E.~C.~G. Sudarshan.
\newblock ``Equivalence of semiclassical and quantum mechanical descriptions of
  statistical light beams''.
\newblock \href{https://dx.doi.org/10.1103/PhysRevLett.10.277}{Physical Review
  Letters {\bf 10}, 277--279}~(1963).

\bibitem{Glauber1963}
Roy~J. Glauber.
\newblock ``Coherent and incoherent states of the radiation field''.
\newblock \href{https://dx.doi.org/10.1103/PhysRev.131.2766}{Physical Review
  {\bf 131}, 2766--2788}~(1963).

\bibitem{Eberlyetal1980}
J.~H. Eberly, N.~B. Narozhny, and J.~J. Sanchez-Mondragon.
\newblock ``Periodic spontaneous collapse and revival in a simple quantum
  model''.
\newblock \href{https://dx.doi.org/10.1103/PhysRevLett.44.1323}{Physical Review
  Letters {\bf 44}, 1323--1326}~(1980).

\bibitem{Rempeetal1987}
Gerhard Rempe, Herbert Walther, and Norbert Klein.
\newblock ``Observation of quantum collapse and revival in a one-atom maser''.
\newblock \href{https://dx.doi.org/10.1103/PhysRevLett.58.353}{Physical Review
  Letters {\bf 58}, 353--356}~(1987).

\bibitem{GeaBanacloche1990}
Julio Gea-Banacloche.
\newblock ``Collapse and revival of the state vector in the {Jaynes-Cummings}
  model: An example of state preparation by a quantum apparatus''.
\newblock \href{https://dx.doi.org/10.1103/PhysRevLett.65.3385}{Physical Review
  Letters {\bf 65}, 3385--3388}~(1990).

\bibitem{GeaBanacloche1991}
Julio Gea-Banacloche.
\newblock ``Atom- and field-state evolution in the {Jaynes-Cummings} model for
  large initial fields''.
\newblock \href{https://dx.doi.org/10.1103/PhysRevA.44.5913}{Physical Review A
  {\bf 44}, 5913--5931}~(1991).

\bibitem{PhoenixKnight1991a}
Simon J.~D. Phoenix and P.~L. Knight.
\newblock ``Establishment of an entangled atom-field state in the
  {Jaynes-Cummings} model''.
\newblock \href{https://dx.doi.org/10.1103/PhysRevA.44.6023}{Physical Review A
  {\bf 44}, 6023--6029}~(1991).

\bibitem{GeaBanacloche2002}
Julio Gea-Banacloche.
\newblock ``Some implications of the quantum nature of laser fields for quantum
  computations''.
\newblock \href{https://dx.doi.org/10.1103/PhysRevA.65.022308}{Physical Review
  A {\bf 65}, 022308}~(2002).

\bibitem{vanEnkKimble2002}
S.J. van Enk and H.J. Kimble.
\newblock ``On the classical character of control fields in quantum information
  processing''.
\newblock \href{https://dx.doi.org/10.26421/QIC2.1}{Quantum Information \&
  Computation {\bf 2}, 1--13}~(2002).

\bibitem{SilberfarbDeutsch}
Andrew Silberfarb and Ivan~H. Deutsch.
\newblock ``Entanglement generated between a single atom and a laser pulse''.
\newblock \href{https://dx.doi.org/10.1103/PhysRevA.69.042308}{Physical Review
  A {\bf 69}, 042308}~(2004).

\bibitem{Raimondetal2001}
J.~M. Raimond, M.~Brune, and S.~Haroche.
\newblock ``Manipulating quantum entanglement with atoms and photons in a
  cavity''.
\newblock \href{https://dx.doi.org/10.1103/RevModPhys.73.565}{Rev. Mod. Phys.
  {\bf 73}, 565--582}~(2001).

\bibitem{Finketal2008}
J.~M. Fink, M.~Göppl, M.~Baur, R.~Bianchetti, P.~J. Leek, A.~Blais, and
  A.~Wallraff.
\newblock ``Climbing the {Jaynes-Cummings} ladder and observing its
  nonlinearity in a cavity {QED} system''.
\newblock \href{https://dx.doi.org/10.1038/nature07112}{Nature {\bf 454},
  315--318}~(2008).

\bibitem{GutierrezJaureguiAgarwal2021}
R.~Guti\'errez-J\'auregui and G.~S. Agarwal.
\newblock ``Probing the spectrum of the {Jaynes-Cummings-Rabi} model by its
  isomorphism to an atom inside a parametric amplifier cavity''.
\newblock \href{https://dx.doi.org/10.1103/PhysRevA.103.023714}{Physical Review
  A {\bf 103}, 023714}~(2021).

\bibitem{Liuetal2021constructing}
Yuan Liu, Jasmine Sinanan-Singh, Matthew~T. Kearney, Gabriel Mintzer, and
  Isaac~L. Chuang.
\newblock ``Constructing qudits from infinite-dimensional oscillators by
  coupling to qubits''.
\newblock \href{https://dx.doi.org/10.1103/PhysRevA.104.032605}{Physical Review
  A {\bf 104}, 032605}~(2021).

\bibitem{Messingeretal2020}
Anette Messinger, Atirach Ritboon, Frances Crimin, Sarah Croke, and Stephen~M
  Barnett.
\newblock ``Coherence and catalysis in the {J}aynes{\textendash}{C}ummings
  model''.
\newblock \href{https://dx.doi.org/10.1088/1367-2630/ab7607}{New Journal of
  Physics {\bf 22}, 043008}~(2020).

\bibitem{KenfackZyczkowski2004}
Anatole Kenfack and Karol Życzkowski.
\newblock ``Negativity of the wigner function as an indicator of
  non-classicality''.
\newblock \href{https://dx.doi.org/10.1088/1464-4266/6/10/003}{Journal of
  Optics B: Quantum and Semiclassical Optics {\bf 6}, 396}~(2004).

\bibitem{Wuetal1986}
Ling-An Wu, H.~J. Kimble, J.~L. Hall, and Huifa Wu.
\newblock ``Generation of squeezed states by parametric down conversion''.
\newblock \href{https://dx.doi.org/10.1103/PhysRevLett.57.2520}{Physical Review
  Letters {\bf 57}, 2520--2523}~(1986).

\bibitem{LoudonKnight1987}
R.~Loudon and P.L. Knight.
\newblock ``Squeezed light''.
\newblock \href{https://dx.doi.org/10.1080/09500348714550721}{Journal of Modern
  Optics {\bf 34}, 709--759}~(1987).

\bibitem{Vahlbruchetal2016}
Henning Vahlbruch, Moritz Mehmet, Karsten Danzmann, and Roman Schnabel.
\newblock ``Detection of 15 {dB} squeezed states of light and their application
  for the absolute calibration of photoelectric quantum efficiency''.
\newblock \href{https://dx.doi.org/10.1103/PhysRevLett.117.110801}{Physical
  Review Letters {\bf 117}, 110801}~(2016).

\bibitem{ScullyZubairy1988}
Marlan~O. Scully and M.S. Zubairy.
\newblock ``Noise free amplification via the two-photon correlated spontaneous
  emission laser''.
\newblock \href{https://dx.doi.org/10.1016/0030-4018(88)90419-1}{Optics
  Communications {\bf 66}, 303--306}~(1988).

\bibitem{TavisCummings1968}
Michael Tavis and Frederick~W. Cummings.
\newblock ``Exact solution for an $n$-molecule---radiation-field
  {H}amiltonian''.
\newblock \href{https://dx.doi.org/10.1103/PhysRev.170.379}{Physical Review
  {\bf 170}, 379--384}~(1968).

\bibitem{Johanssonetal2012}
J.R. Johansson, P.D. Nation, and Franco Nori.
\newblock ``{QuTiP: An open-source Python framework for the dynamics of open
  quantum systems}''.
\newblock \href{https://dx.doi.org/10.1016/j.cpc.2012.02.021}{Computer Physics
  Communications {\bf 183}, 1760--1772}~(2012).

\bibitem{Johanssonetal2013}
J.R. Johansson, P.D. Nation, and Franco Nori.
\newblock ``{QuTiP 2: A Python framework for the dynamics of open quantum
  systems}''.
\newblock \href{https://dx.doi.org/10.1016/j.cpc.2012.11.019}{Computer Physics
  Communications {\bf 184}, 1234--1240}~(2013).

\bibitem{DrobnyJex1993}
Gabriel Drobný and Igor Jex.
\newblock ``The system of {N} two-level atoms interacting with a field mode:
  entanglement and parametric approximation''.
\newblock \href{https://dx.doi.org/10.1016/0030-4018(93)90485-N}{Optics
  Communications {\bf 102}, 141--154}~(1993).

\bibitem{Chumakovetal1994}
S.~M. Chumakov, A.~B. Klimov, and J.~J. Sanchez-Mondragon.
\newblock ``General properties of quantum optical systems in a strong-field
  limit''.
\newblock \href{https://dx.doi.org/10.1103/PhysRevA.49.4972}{Physical Review A
  {\bf 49}, 4972--4978}~(1994).

\bibitem{KlimovChumakov1995}
A.B. Klimov and S.M. Chumakov.
\newblock ``Semiclassical quantization of the evolution operator for a class of
  optical models''.
\newblock \href{https://dx.doi.org/10.1016/0375-9601(95)00342-Z}{Physics
  Letters A {\bf 202}, 145--154}~(1995).

\bibitem{Retamaletal1997}
J.~C. Retamal, C.~Saavedra, A.~B. Klimov, and S.~M. Chumakov.
\newblock ``Squeezing of light by a collection of atoms''.
\newblock \href{https://dx.doi.org/10.1103/PhysRevA.55.2413}{Physical Review A
  {\bf 55}, 2413--2425}~(1997).

\bibitem{PhoenixKnight1991b}
Simon J.~D. Phoenix and P.~L. Knight.
\newblock ``Establishment of an entangled atom-field state in the
  {Jaynes-Cummings} model''.
\newblock \href{https://dx.doi.org/10.1103/PhysRevA.44.6023}{Physical Review A
  {\bf 44}, 6023--6029}~(1991).

\bibitem{GeaBanacloche1992}
J.~Gea-Banacloche.
\newblock ``A new look at the {Jaynes-Cummings} model for large fields: Bloch
  sphere evolution and detuning effects''.
\newblock \href{https://dx.doi.org/10.1016/0030-4018(92)90082-3}{Optics
  Communications {\bf 88}, 531--550}~(1992).

\bibitem{BarnettVaccaro2007}
Stephen~M Barnett and John~A Vaccaro.
\newblock ``The quantum phase operator: A review''.
\newblock \href{https://dx.doi.org/10.1201/b16006}{Taylor \& Francis}. ~(2007).

\bibitem{Scharf1970}
G.~Scharf.
\newblock ``On a quantum mechanical maser model''.
\newblock \href{https://dx.doi.org/10.5169/seals-114194}{Helvetica Physica Acta
  {\bf 43}, 806}~(1970).

\bibitem{HeppLieb1973}
Klaus Hepp and Elliott~H Lieb.
\newblock ``On the superradiant phase transition for molecules in a quantized
  radiation field: the dicke maser model''.
\newblock \href{https://dx.doi.org/10.1016/0003-4916(73)90039-0}{Annals of
  Physics {\bf 76}, 360--404}~(1973).

\bibitem{Bogoliubovetal1996}
N~M Bogoliubov, R~K Bullough, and J~Timonen.
\newblock ``Exact solution of generalized {Tavis - Cummings} models in quantum
  optics''.
\newblock \href{https://dx.doi.org/10.1088/0305-4470/29/19/015}{Journal of
  Physics A: Mathematical and General {\bf 29}, 6305--6312}~(1996).

\bibitem{Luetal2021}
Wangjun Lu, Jie Chen, Le-Man Kuang, and Xiaoguang Wang.
\newblock ``Optimal state for a tavis-cummings quantum battery via the bethe
  ansatz method''.
\newblock \href{https://dx.doi.org/10.1103/PhysRevA.104.043706}{Physical Review
  A {\bf 104}, 043706}~(2021).

\bibitem{LIGO2011}
{The L. I. G. O. Scientific Collaboration}.
\newblock ``A gravitational wave observatory operating beyond the quantum
  shot-noise limit''.
\newblock \href{https://dx.doi.org/10.1038/nphys2083}{Nature Physics {\bf 7},
  962}~(2011).

\bibitem{Madsenetal2022}
Lars~S. Madsen, Fabian Laudenbach, Mohsen~Falamarzi. Askarani, Fabien Rortais,
  Trevor Vincent, Jacob F.~F. Bulmer, Filippo~M. Miatto, Leonhard Neuhaus,
  Lukas~G. Helt, Matthew~J. Collins, Adriana~E. Lita, Thomas Gerrits, Sae~Woo
  Nam, Varun~D. Vaidya, Matteo Menotti, Ish Dhand, Zachary Vernon, Nicolás
  Quesada, and Jonathan Lavoie.
\newblock ``Quantum computational advantage with a programmable photonic
  processor''.
\newblock \href{https://dx.doi.org/10.1038/s41586-022-04725-x}{Nature {\bf
  606}, 75--81}~(2022).

\bibitem{KiilerichMolmer2019}
Alexander~Holm Kiilerich and Klaus M\o{}lmer.
\newblock ``Input-output theory with quantum pulses''.
\newblock \href{https://dx.doi.org/10.1103/PhysRevLett.123.123604}{Physical
  Review Letters {\bf 123}, 123604}~(2019).

\bibitem{KiilerichMolmer2020}
Alexander~Holm Kiilerich and Klaus M\o{}lmer.
\newblock ``Quantum interactions with pulses of radiation''.
\newblock \href{https://dx.doi.org/10.1103/PhysRevA.102.023717}{Physical Review
  A {\bf 102}, 023717}~(2020).

\end{thebibliography}

\onecolumn\newpage
\appendix

\section{Averaging fidelity of $\tfrac{\pi}{2}$ pulses over $\theta,\phi$}~
\label{app:averaging fidelity}

The overlap between the state $\ket{\Psi(t)}$ given in Eq. \eqref{eq:transformed joint state from arbitrary initial} and the desired evolved state
$\ket{\theta,\phi}_{\frac{\pi}{2}}$ is
\eq{
\vphantom{a}_{\frac{\pi}{2}}\braket{\theta,\phi}{\Psi(t)}
&=\sum_n\ket{n}\left[
\bra{\mathrm{g}}\left(\ket{G_n}\cos\frac{\theta}{2}+\ket{E_n}\sin\frac{\theta}{2}\eu^{\iu\phi}\right)\frac{\cos\frac{\theta}{2}-\eu^{-\iu\phi}\sin\frac{\theta}{2}}{\sqrt{2}}+
%%%
\right.\\
&\qquad\left.
\bra{\mathrm{e}}\left(\ket{G_n}\cos\frac{\theta}{2}+\ket{E_n}\sin\frac{\theta}{2}\eu^{\iu\phi}\right)\frac{\cos\frac{\theta}{2}+\eu^{-\iu\phi}\sin\frac{\theta}{2}}{\sqrt{2}}
\right].
}
The norm of this state provides the fidelity between $\ket{\theta,\phi}_{\frac{\pi}{2}}$ and the  reduced density matrix corresponding to the evolved atomic system. If we average this result over all possible initial values of $\phi$, all of the $\phi$-dependent terms vanish. The pertinent integrals over all possible initial values of $\theta$ yield either $1/3$ or $2/3$, becoming:
\eq{
\mathcal{F}\left(\left\{\psi_n\right\},t\right)&=\frac{1}{4\pi}\int_0^{2\pi}d\phi\int_0^\pi\sin\theta \,d\theta\,\left|\vphantom{a}_{\frac{\pi}{2}}\braket{\theta,\phi}{\Psi(t)}\right|^2\\
&=\sum_n\frac{1}{4}\left(\left|\braket{g}{G_n}\right|^2+\left|\braket{e}{G_n}\right|^2+\left|\braket{g}{E_n}\right|^2+\left|\braket{e}{E_n}\right|^2\right)
+\frac{1}{6}\RE\left(
\braket{e}{E_n}\braket{G_n}{e}
\right.\\
-
\braket{e}{E_n}&\left.\braket{E_n}{g}
-
\braket{e}{G_n}\braket{E_n}{g}
+\braket{e}{E_n}\braket{G_n}{g}-
\braket{g}{E_n}\braket{G_n}{g}
+
\braket{e}{G_n}\braket{G_n}{g}
\right).
}
We know from normalization of $\ket{\Psi}$ that $\sum_n\left|\braket{g}{G_n}\right|^2+\left|\braket{e}{G_n}\right|^2=\sum_n\left|\braket{g}{E_n}\right|^2+\left|\braket{e}{E_n}\right|^2=1$, and we can readily compute the remaining terms:
\eq{
\mathcal{F}=&\frac{1}{2}+\frac{1}{6}\sum_n\RE\left[\left|\psi_n\right|^2 \cos\frac{\Omega_n t}{2}\cos\frac{\Omega_{n-1}t}{2}-
\psi_{n-1}\psi_{n+1}^*\sin\frac{\Omega_n t}{2}\sin\frac{\Omega_{n-1}t}{2}
\right.\\
&\left.+\iu \psi_n \psi_{n+1}^*\sin\frac{\Omega_{n}t}{2}\left(
2\cos\frac{\Omega_n t}{2}+
\cos\frac{\Omega_{n-1}t}{2}+
\cos\frac{\Omega_{n+1}t}{2}
\right)
\right].
}

We can optimize this averaged fidelity by specifying a fixed phase relationship between neighbouring photon-number states: \eq{\arg \psi_{n+1}=\frac{\pi}{2}+\arg \psi_n.} Then, the averaged fidelity becomes
\eq{
\mathcal{F}=&\frac{1}{2}+\frac{1}{6}\sum_n\left|\psi_n\right|^2 \cos\frac{\Omega_n t}{2}\cos\frac{\Omega_{n-1}t}{2}+\left|
\psi_{n-1}\psi_{n+1}\right|\sin\frac{\Omega_n t}{2}\sin\frac{\Omega_{n-1}t}{2}
\\
&+ \left|\psi_n \psi_{n+1}\right|\sin\frac{\Omega_{n}t}{2}\left(
2\cos\frac{\Omega_n t}{2}+
\cos\frac{\Omega_{n-1}t}{2}+
\cos\frac{\Omega_{n+1}t}{2}
\right)
} and we can seek the photon-number distribution $\{|\psi_n|^2\}$ that optimizes this quantity.

\section{Averaging fidelity over $\theta,\phi$ for any rotation}~
\label{app:averaging fidelity any pulse}
We now consider the overlap
 between the state $\ket{\Psi(t)}$ given in Eq. \eqref{eq:transformed joint state from arbitrary initial} and the desired evolved state
\eq{
\ket{\theta,\phi}_{\Theta}=\eu^{\iu\Theta \frac{\sigma_y}{2}}\ket{\theta,\phi}=\left(\cos\frac{\theta}{2}\cos\frac{\Theta}{2}-\eu^{\iu\phi}\sin\frac{\theta}{2}\sin\frac{\Theta}{2}\right)\ket{\mathrm{g}}+
\left(\cos\frac{\theta}{2}\sin\frac{\Theta}{2}+\eu^{\iu\phi}\sin\frac{\theta}{2}\cos\frac{\Theta}{2}\right)\ket{\mathrm{e}}.
}
The overlap is given by
\eq{
\vphantom{a}_{\Theta}\braket{\theta,\phi}{\Psi(t)}
&=\sum_n\ket{n}\left[
\bra{\mathrm{g}}\left(\ket{G_n}\cos\frac{\theta}{2}+\ket{E_n}\sin\frac{\theta}{2}\eu^{\iu\phi}\right)\left(\cos\frac{\theta}{2}\cos\frac{\Theta}{2}-\eu^{-\iu\phi}\sin\frac{\theta}{2}\sin\frac{\Theta}{2}\right)
\right.\\
&\qquad\qquad+\left.
%%%
\bra{\mathrm{e}}\left(\ket{G_n}\cos\frac{\theta}{2}+\ket{E_n}\sin\frac{\theta}{2}\eu^{\iu\phi}\right)
\left(\cos\frac{\theta}{2}\sin\frac{\Theta}{2}+\eu^{-\iu\phi}\sin\frac{\theta}{2}\cos\frac{\Theta}{2}\right)
\right].
}
The norm of this state provides the fidelity between $\ket{\theta,\phi}_{\Theta}$ and the  reduced density matrix corresponding to the evolved atomic system. If we average this result over all possible initial values of $\phi$, all of the $\phi$-dependent terms vanish. The pertinent integrals over all possible initial values of $\theta$ can be explicitly computed, becoming:
\eq{
\mathcal{F}&\left(\left\{\psi_n\right\},t\right)=\frac{1}{4\pi}\int_0^{2\pi}d\phi\int_0^\pi\sin\theta \,d\theta\,\left|\vphantom{a}_{\Theta}\braket{\theta,\phi}{\Psi(t)}\right|^2\\
&\qquad\qquad=\sum_n\frac{1}{4}\left(\left|\braket{g}{G_n}\right|^2+\left|\braket{e}{G_n}\right|^2+\left|\braket{g}{E_n}\right|^2+\left|\braket{e}{E_n}\right|^2\right)
\\
&+
%%%%%%%%%
\frac{\sin \Theta}{12}  \left[\left(\braket{E_n}{e}+\braket{G_n}{g}\right) \left(\braket{e}{G_n}-\braket{g}{E_n}\right)-\left(\braket{e}{E_n}+\braket{g}{G_n}\right) \left(\braket{E_n}{g}-\braket{G_n}{e}\right)\right]
\\&+
\frac{\cos \Theta}{12} \left[\left(\braket{e}{E_n}+\braket{g}{G_n}\right) \left(\braket{E_n}{e}+\braket{G_n}{g}\right)+\left(\braket{e}{G_n}-\braket{g}{E_n}\right) \left(\braket{E_n}{g}-\braket{G_n}{e}\right)\right]
\\
&+\frac{1}{12}\left(\braket{e}{E_n} \braket{G_n}{g}+\braket{E_n}{e} \braket{g}{G_n}-\braket{e}{G_n} \braket{E_n}{g}-\braket{g}{E_n} \braket{G_n}{e}\right)
.
}
We know from normalization of $\ket{\Psi}$ that $\sum_n\left|\braket{g}{G_n}\right|^2+\left|\braket{e}{G_n}\right|^2=\sum_n\left|\braket{g}{E_n}\right|^2+\left|\braket{e}{E_n}\right|^2=1$, and we can readily compute the remaining terms:
\eq{
\mathcal{F}=\frac{1}{2}
+\sum_n &\left|\psi_n\right|^2\left(\frac{1}{3}\cos\frac{\Omega_n t}{2}\cos\frac{\Omega_{n-1} t}{2}\cos^2\frac{\Theta}{2}+\frac{\cos\Omega_n t+\cos\Omega_{n-1}t}{12}\cos\Theta\right)
\\&-\frac{1}{3}\RE\left(\psi_{n-1}\psi_{n+1}^*\right)\sin\frac{\Omega_n t}{2}\sin\frac{\Omega_{n-1} t}{2}\sin^2\frac{\Theta}{2}\\
&+\frac{1}{6}\left(\cos\frac{\Omega_n t}{2}+
\cos\frac{\Omega_{n-1} t}{2}\right)
\left[\IM\left(\psi_n^*\psi_{n+1}\right)\sin\frac{\Omega_n t}{2}-\IM\left(\psi_n^*\psi_{n-1}\right)
\sin\frac{\Omega_{n-1} t}{2}\right]\sin\Theta
.
}

%In the relevant region $\Omega_{n}t \lesssim 2\pi$, the term $\cos\frac{\Omega_n t}{2}-
%\cos\frac{\Omega_{n-1} t}{2}$ is close to zero and is strictly negative. In contrast, all of $\sin\frac{\Omega_n t}{2}-
%\sin\frac{\Omega_{n-1} t}{2}$, $\sin\frac{\Omega_n t}{2}$, and $
%\sin\frac{\Omega_{n-1} t}{2}$ are positive.
We can optimize the averaged fidelity by specifying a fixed phase relationship between neighbouring photon-number states: \eq{\arg \psi_{n+1}=\frac{\pi}{2}+\arg \psi_n.} Then, the averaged fidelity becomes
\eq{
\mathcal{F}\approx \frac{1}{2}
+\sum_n &\left|\psi_n\right|^2\left(\frac{1}{3}\cos\frac{\Omega_n t}{2}\cos\frac{\Omega_{n-1} t}{2}\cos^2\frac{\Theta}{2}+\frac{\cos\Omega_n t+\cos\Omega_{n-1}t}{12}\cos\Theta\right)\\
&
+\frac{1}{3}\left|\psi_{n-1}\psi_{n+1}\right|\sin\frac{\Omega_n t}{2}\sin\frac{\Omega_{n-1} t}{2}\sin^2\frac{\Theta}{2}\\
&+\frac{1}{6}\left(\cos\frac{\Omega_{n-1} t}{2}+
\cos\frac{\Omega_{n} t}{2}\right)
\left(\left|\psi_n \psi_{n+1}\right|\sin\frac{\Omega_n t}{2}+\left|\psi_{n-1}\psi_n\right|
\sin\frac{\Omega_{n-1} t}{2}\right)\sin\Theta
} and we can seek the photon-number distribution $\{|\psi_n|^2\}$ that optimizes this quantity. The opposite phase relationship is necessary when the sign of $\sin\Theta$ changes.

\section{Averaging fidelity over $\theta$ for known $\phi$ and any rotation}~
\label{app:averaging fidelity any pulse fixed phi}
We next compute the average fidelity for a known $\phi$: this is physically equivalent to setting $\phi=0$. Then, we must extend the $\theta$ coordinate to range from $0$ to $2\pi$ and remember that the relevant Jacobian is now unity.

The averaged fidelity now becomes
\eq{
\mathcal{F}&\left(\left\{\psi_n\right\},t\right)=\frac{1}{2\pi}\int_0^{2\pi} \,d\theta\,\left|\vphantom{a}_{\Theta}\braket{\theta,0}{\Psi(t)}\right|^2
\\
&\qquad\qquad=\sum_n\frac{1}{4}\left(\left|\braket{g}{G_n}\right|^2+\left|\braket{e}{G_n}\right|^2+\left|\braket{g}{E_n}\right|^2+\left|\braket{e}{E_n}\right|^2\right)
\\
&+
%%%%%%%%%
\frac{\sin \Theta}{8}  \left[\left(\braket{E_n}{e}+\braket{G_n}{g}\right) \left(\braket{e}{G_n}-\braket{g}{E_n}\right)-\left(\braket{e}{E_n}+\braket{g}{G_n}\right) \left(\braket{E_n}{g}-\braket{G_n}{e}\right)\right]
\\&+
\frac{\cos \Theta}{8} \left[\left(\braket{e}{E_n}+\braket{g}{G_n}\right) \left(\braket{E_n}{e}+\braket{G_n}{g}\right)+\left(\braket{e}{G_n}-\braket{g}{E_n}\right) \left(\braket{E_n}{g}-\braket{G_n}{e}\right)\right]
.
}
We know from normalization of $\ket{\Psi}$ that $\sum_n\left|\braket{g}{G_n}\right|^2+\left|\braket{e}{G_n}\right|^2=\sum_n\left|\braket{g}{E_n}\right|^2+\left|\braket{e}{E_n}\right|^2=1$, and we can readily compute the remaining terms:
\eq{
\mathcal{F}=\frac{1}{2}
+\sum_n &\left|\psi_n\right|^2\left(\frac{1}{4}\cos\frac{\Omega_n t}{2}\cos\frac{\Omega_{n-1} t}{2}+\frac{\cos\Omega_n t+\cos\Omega_{n-1}t}{8}\right)\cos\Theta
\\&+\frac{1}{4}\RE\left(\psi_{n-1}\psi_{n+1}^*\right)\sin\frac{\Omega_n t}{2}\sin\frac{\Omega_{n-1} t}{2}\cos\Theta\\
&+\frac{1}{4}\left(\cos\frac{\Omega_n t}{2}+
\cos\frac{\Omega_{n-1} t}{2}\right)
\left[\IM\left(\psi_n^*\psi_{n+1}\right)\sin\frac{\Omega_n t}{2}-\IM\left(\psi_n^*\psi_{n-1}\right)
\sin\frac{\Omega_{n-1} t}{2}\right]\sin\Theta
.
}

It is now not as easy to choose the phase relationship for the coefficients. The region $\tfrac{\pi}{2}\leq \Theta\leq \pi$ is optimized by the usual 
\eq{\arg \psi_{n+1}=\frac{\pi}{2}+\arg \psi_n,} because $\cos\Theta \leq 0$ and $\RE\left(\psi_{n-1}\psi_{n+1}^*\right)\leq 0$, which achieves
\eq{
\mathcal{F}=\frac{1}{2}
+\sum_n &\left|\psi_n\right|^2\left(\frac{1}{4}\cos\frac{\Omega_n t}{2}\cos\frac{\Omega_{n-1} t}{2}+\frac{\cos\Omega_n t+\cos\Omega_{n-1}t}{8}\right)\cos\Theta
\\&-\frac{1}{4}\left|\psi_{n-1}\psi_{n+1}\right|\sin\frac{\Omega_n t}{2}\sin\frac{\Omega_{n-1} t}{2}\cos\Theta\\
&+\frac{1}{4}\left|\psi_n \psi_{n+1}\right|\sin\frac{\Omega_n t}{2}\left(2\cos\frac{\Omega_n t}{2}+
\cos\frac{\Omega_{n-1} t}{2}+
\cos\frac{\Omega_{n+1} t}{2}\right)\sin\Theta
.
} The same can be done in all regions where the signs of $\cos\Theta$ and $\sin\Theta$ differ, but otherwise we cannot guarantee all terms in the sum to be positive. Instead, one is led to a trade off that must be optimized for each value of $\Theta$ in turn.

Restricting our attention to $\tfrac{\pi}{2}$ pulses, we find the averaged fidelity
\eq{
\mathcal{F}=\frac{1}{2}
+\frac{1}{4}\sum_n \left|\psi_n \psi_{n+1}\right|\sin\frac{\Omega_n t}{2}\left(2\cos\frac{\Omega_n t}{2}+
\cos\frac{\Omega_{n-1} t}{2}+
\cos\frac{\Omega_{n+1} t}{2}\right)
.
}
If we instead leave arbitrary the phase difference \eq{
\varphi\equiv \arg \psi_{n+1}-\arg \psi_{n},
} we arrive at the averaged fidelity
\eq{
\mathcal{F}=\frac{1}{2}
+\sum_n &\left|\psi_n\right|^2\left(\frac{1}{4}\cos\frac{\Omega_n t}{2}\cos\frac{\Omega_{n-1} t}{2}+\frac{\cos\Omega_n t+\cos\Omega_{n-1}t}{8}\right)\cos\Theta
\\&+\frac{\cos(2\varphi)}{4}\left|\psi_{n-1}\psi_{n+1}\right|\sin\frac{\Omega_n t}{2}\sin\frac{\Omega_{n-1} t}{2}\cos\Theta\\
&+\frac{\sin\varphi}{4}\left|\psi_n \psi_{n+1}\right|\sin\frac{\Omega_n t}{2}\left(2\cos\frac{\Omega_n t}{2}+
\cos\frac{\Omega_{n-1} t}{2}+
\cos\frac{\Omega_{n+1} t}{2}\right)\sin\Theta
.
}

\end{document}